\documentclass[12pt]{iopart}
\usepackage{iopams}
\usepackage{setstack}
\usepackage{lmodern} 
\usepackage{graphicx}
\usepackage{epigraph}
\usepackage{amssymb}
\usepackage{paralist}
\usepackage{color}
\usepackage{xfrac} 
\usepackage{tikz}
\usepackage{adjustbox}

\usepackage{amscd}
\usepackage{dsfont}
\usepackage{bm}

\newif\ifdraft
\draftfalse

\DeclareGraphicsExtensions{.pdf, .png, .bmp, .jpg}

\newcommand{\operatorname}[1]{\text{#1}}
\newcommand{\eqref}[1]{(\ref{#1})}

\begin{document}

\title[Spin Fermion Mapping]{A Mapping between the Spin and Fermion Algebra}

\author{Felix Meier, Daniel Waltner, Petr Braun\footnote[1]{Deceased.}, Thomas Guhr}
\address{
Faculty of Physics, University Duisburg-Essen, Lotharstr. 1, 47048 Duisburg, Germany;
}
\begin{abstract}
We derive a formalism to express the spin algebra $\mathfrak{su}(2)$ in a spin $s$ representation in terms of the algebra of $L$ fermionic operators that obey the Canonical Anti-commutation Relations.
We also give the reverse direction of expressing the fermionic operators as polynomials in the spin operators of a single spin.
We extend here to further spin values the previous investigations by Dobrov [J.Phys.A: Math. Gen. 36, L503 (2003)] who in turn clarified on an inconsistency within a similar formalism in the works of Batista and Ortiz [Phys.\ Rev.\ Lett. 86, 1082 (2001)].
We then consider a system of $L$ fermion flavors and apply our mapping in order to express it in terms of the spin algebra.
Furthermore we investigate a possibility to simplify certain Hamiltonian operators by means of the mapping.
\end{abstract}


Keywords: Fermions, Spin Algebra, Jordan-Wigner transformation
\maketitle
\section{Introduction}
While it has never been easier to harvest the power of numerical simulations of a given physical system of interest, there still is a lot of interest in analytically treatable models as well.
One important tool for analytical calculations in the context of lattice systems is the Jordan-Wigner transformation \cite{JWT}, which relates a spin chain for spin $s=1/2$ to a system of spinless fermions.
Approaches to generalize this transformation to higher spin $s$ have been conducted e.g. in \cite{General_JWT_Batista},\cite{General_JWT_Kochmanski}, \cite{General_JWT_Galitski} and \cite{Dobrov} though in \cite{Dobrov} it is shown, that the mapping from \cite{General_JWT_Batista} yields an inconsistency.
Furthermore in \cite{Dobrov} a spin in the representation $s=3/2$ is expressed by two fermionic operators by a direct case-specific calculation, which we generalize to a complete formalism to do the same for arbitrary numbers of fermions.
Different to the Jordan-Wigner transformation we obtain a mapping between $L$ fermionic operators and a single spin of a suitably chosen representation $s$, as opposed to a whole chain of spins.
We want to stress that our mapping has no direct mathematical connection with the Jordan-Wigner transformation and in the form presented here is not applicable to general spin chain systems, though it may be possible to extend it to the latter types of systems in a similar vein as described in \cite{Dobrov}.
The topical similarity with the Jordan-Wigner transformation is that it is also used for obtaining exact simplifications of complicated Hamiltonians.
Works like \cite{minami5}, \cite{minami6} contain formalisms directly applied to certain spin chain systems including cases beyond spin $1/2$, which suffer from the same downside as our methodology, namely that explicit expressions for higher spins quickly become too difficult to deal with even if they are constituted on an analytical approach.
\\
\\
We assume a number of $L$ fermion flavors, so that the fermionic creation-/annihilation-operators $\hat{c}^\dagger_\alpha,\hat{c}_\alpha$ can be labeled by an integer index $\alpha$ which ranges from 1 to $L$.
Since on an abstract level the algebras under study are not isomorphic, our mapping can only exist inside of a representation.
By then picking a suitable representation of both the Canonical Anti-commutation Relations (CAR), defining the fermion algebra and $\mathfrak{su}(2)$, defining the spin algebra, it is finally possible to map spin operators onto combinations of fermion operators and vice versa.
In a sense, one could also view our spin fermion mapping as an alternative take on the usual coupling theory \cite{Guhr_quelle}, \cite{CLEBSCH_GORDAN} for representing a large spin representation in terms of smaller ones. 
The usual formalism coined addition of angular momenta and spins involves the computation of Clebsch-Gordan coefficients or more generally the computation of $3$-$j$, $6$-$j$ and $9$-$j$-symbols \cite{Guhr_quelle}.
Here we have a collection of fermionic operators in the place of the lower spin representations which are coupled to form the algebra of a spin $s$ representation by taking certain linear combinations of products of fermion operators.
The inverse mapping of fermionic operators onto spins turns out to be non-injective, since it involves the computation of a rational power of a matrix.
The mapping is first derived in the form of a recursion formula that is the primary tool also for practical calculations, which later on is given a closed form solution as well.
Due to the recursivity, the first formulation is also well suited for automated symbolic computations, while the non-recursive version, due to being quite cumbersome, is only of theoretical interest, namely regarding its existence.
One can choose the spin representation of interest and then exactly compute the expansion coefficients in terms of fermionic operators with our formalism.
Approaches like the one presented in this paper are also of interest in the context of chemical physics \cite{Miller1}, \cite{Miller2}, \cite{Miller3}.
Please note, that in all of this work we are dealing with concrete matrix representations of the algebras under consideration, so that there never occur any operators defined only abstractly.
\\
\\
The paper is organized as follows: in section two we first introduce the general formalism based on expressing spins by fermions and derive aforementioned recursion relation.
In the second part of section two we give the mapping in the inverse direction, expressing fermions by spins.
For the operator $\hat{S}_z$ we also get a very compact analytical expression in terms of fermionic number operators, as well as an inverse relation for fermionic number operators in terms of polynomials of $\hat{S}_z$.
Finally we conclude section two with a closed form solution of the recursion relation.
In section three we consider explicit examples of practical applications of the formalism.
We give an example of a specific non-diagonal Hamiltonian cubic in the fermion operators and show, that it corresponds to a simple Hamiltonian linear in the spin operators.
Here we show, how the spin fermion mapping can be applied to diagonalize fermionic Hamiltonians and we end with an example of rewriting an Ising interaction term for spin $3/2$ in terms of number operators.
Finally, we supply various proofs of relations used in the main text as well as several explicit results of our formalism in the appendix.

\section{Deriving the Spin Fermion Mapping}
\subsection{Fermionic Operators for a System of $L$ fermionic Flavors}
The fermionic CAR are defined as
\begin{equation}\label{CAR_1}
	\{\hat{c},\hat{c}^\dagger\}=\mathds{1}_2\,\,\,\text{,}\\
	\hat{c}:=\left[\begin{array}
			[c]{cc}%
			0 & 0\\
			1 & 0%
		\end{array}\right]\hspace{2cm}
	\hat{c}^\dagger:=\left[\begin{array}
					[c]{cc}%
					0 & 1\\
					0 & 0%
				\end{array}\right]
\end{equation}
where $\hat{c}$ and $\hat{c}^\dagger$,with the latter denoting the complex conjugate transpose of the former, are referred to as the fermionic operators and $\mathds{1}_2$ is the two-dimensional identity matrix.
The generalization onto a system of $L$ operators fulfilling
\begin{equation}\label{CAR}
	\{\hat{c}_\alpha,\hat{c}^\dagger_\beta\}=\delta_{\alpha\beta}\mathds{1}_{2^L},\,\,\,\alpha,\beta\in\{1,\cdots,L\}
\end{equation}
can be realized (among other choices amounting to a different ordering of the tensor factors) by the matrices 
\begin{equation}\label{c_alpha}
	\hat{c}_\alpha:=\sigma_z^{\otimes (\alpha -1)} \otimes \hat{c} \otimes 
\mathds{1}_2^{\otimes (L-\alpha)}\in \operatorname{Mat}(2^L,2^L)
\end{equation}
\begin{eqnarray}\label{c_alpha_dagger}
	&\hat{c}^\dagger_\alpha:=\sigma_z^{\otimes (\alpha -1)} \otimes \hat{c}^\dagger \otimes \mathds{1}_2^{\otimes (L-\alpha)} \in \operatorname{Mat}(2^L,2^L)
\end{eqnarray}
as proven in the appendix (\ref{A.1.}).
Note, that the usual two-dimensional fermion operators $\hat{c},\hat{c}^\dagger$ are a special case of the $2^L$-dimensional ones and are obtained for $L=1$ as $\hat{c}=\hat{c}_1, \hat{c}^\dagger=\hat{c}_1^\dagger$, where the index $\alpha$ only takes this single value $1$.
We also define the number operators for site $\alpha$:
\begin{equation}\label{n_alpha}
	\hat{n}_\alpha:=\hat{c}_\alpha^\dagger 
\hat{c}_\alpha=\mathds{1}_2^{\otimes(\alpha-1)}\otimes \hat{c}^\dagger \hat{c} 
\otimes \mathds{1}_{2}^{\otimes(L-\alpha)}\in \operatorname{Mat}(2^L,2^L)\,\,\,\text{.}
	\end{equation}
On the other hand the spin algebra in terms of the operators $\hat{S}_x$, $\hat{S}_y$, $\hat{S}_z$, or the common alternative choice of the set of operators $\hat{S}_z$, $\hat{S}_\pm:=\hat{S}_x\pm i\hat{S}_y$ can be defined by the commutation relations \cite{CLEBSCH_GORDAN}
\begin{eqnarray}\label{su2}
	[\hat{S}_z,\hat{S}_\pm]=\pm\hat{S}_\pm\hspace{2cm} 
[\hat{S}_+,\hat{S}_-]=2\hat{S}_z\,\,\,\text{,}\\
	\hat{S}_{j}\in\operatorname{Mat}(2s+1,2s+1)\nonumber
\end{eqnarray}
which, together with $\hat{S}_+^\dagger=\hat{S}_-$, shows, that one needs only one of the spin ladder operators $\hat{S}_\pm$ to reconstruct all of the other operators.
While the most precise and logically accurate definition of the spin operators as matrix representations of abstract generators $\hat{T}_j$ of the Lie Algebra $\mathfrak{su}(2)$ would be by means of a Lie Algebra representation $\rho^{(s)}$ for spin $s$ as
\begin{eqnarray}
	&\rho^{(s)} : \mathfrak{su}(2)\rightarrow \operatorname{Mat}(2s+1,2s+1)\\
	\,\,\,\,& \hat{T}_j \mapsto \rho^{(s)}(\hat{T}_j):=\hat{S}^{(s)}_j,\nonumber
\end{eqnarray}
we use the usual convention of not denoting the explicit superscript $(s)$ and without explicit mention of the representation map $\rho^{(s)}$. Instead we just assume that the operators are given in terms of the matrices $\hat{S}_j$ for the spin $s$ representation, so that they fulfill the commutation relations (\ref{su2}).
Similarly we do not denote the fermion operators for $L$ flavors as $\hat{c}_\alpha (L)$, but as above (\ref{c_alpha}),(\ref{c_alpha_dagger}), simply as $\hat{c}_\alpha$ and $\hat{c}^\dagger_\alpha$ in order to facilitate the notation.
The parameters $s$, $L$ are kept undetermined outside of explicit examples, where they are given concrete numerical values in the accompanying text.
We think of $\hat{S}_z, \hat{S}_\pm$ as the spin-operators of a fictitious particle with spin $s$ such that\cite{Dobrov}
\begin{equation}\label{compatibility}
	2s+1=2^L=:N
\end{equation}
Since the right hand side is always even, only half-integer spin-representations $s=q/2$, with odd numbers $q\in\mathbb{N}$ make an appearance.
Our aim is to show, how they can be expressed in terms of the matrices $\hat{c}_\alpha , \hat{c}_\alpha^\dagger$ and, vice versa, how to express $\hat{c}_\alpha, \hat{c}^\dagger_\alpha$ in terms of powers of $\hat{S}_z$ and $\hat{S}_\pm$.
It will be convenient to introduce $N\times N$ matrices $\tilde{E}_{j}$ with a single non-zero element $(\tilde{E}_{j})_{j\,\,j+1}=1$.
Any matrix with its only non-zero entries in the first upper off-diagonal can be expressed as a linear combination of $\{\tilde{E}_{j}\}_{k\in \{1,\cdots ,N-1\}}$, e.g.,
\begin{equation}\label{Sp_components}
	\hat{S}_+=\sum\limits_{j=1}^{N-1}\sqrt{j(2s+1-j)}\tilde{E}_{j}\,\,\,\text{,}
\end{equation}
where the components $(\hat{S}_+)_{jk}$ are indexed by $j,k\in\{1,\cdots,2s+1\}$ instead of the more conventional $j,k\in\{s,\cdots, -s\}$, which changes the usual form of the components $(\hat{S}_+)_{j\,\, j+1}=\sqrt{(s-j)(s+j+1)}$ \cite{CLEBSCH_GORDAN} to the ones above.
For matrices $A$ with components $A_{jk}$ we use the notation $A=[A_{jk}]$ as well as $(A)_{jk}=A_{jk}$ and analogously for vectors viewed as matrices with just one column.

\subsection{Mapping Spins onto Fermions}
The fermion operators $\hat{c}_\alpha^\dagger$ (\ref{c_alpha_dagger}) can all be written as a power $M_\alpha^{2^{L-\alpha}}$ of an upper off-diagonal matrix (UODM) $M_\alpha$, meaning a matrix which has its only non-zero entries on its first upper off-diagonal, as one can see for example for $L=2$:
\begin{equation}\label{root_of_fermion}
	 \hat{c}^\dagger_1=\hat{c}^\dagger\otimes\mathds{1}_2=\left[\begin{array}{cccc}
		0 & 0 & 1 & 0\\
		0 & 0 & 0 & 1\\
		0 & 0 & 0 & 0\\
		0 & 0 & 0 & 0
	\end{array}\right]
	 =\left[\begin{array}{cccc}
		0 & \pm1 & 0 & 0\\
		0 & 0 & \pm1 & 0\\
		0 & 0 & 0 & \pm1\\
		0 & 0 & 0 & 0
	\end{array}\right]^2
\end{equation}
\begin{equation}
	\hat{c}^\dagger_2=\sigma_z\otimes \hat{c}^\dagger=\left[\begin{array}{cccc}
		0 & 1 & 0 & 0\\
		0 & 0 & 0 & 0\\
		0 & 0 & 0 & -1\\
		0 & 0 & 0 & 0
	\end{array}\right]\,\,\,\text{,}
\end{equation}
where the $\pm$ are meant to be either all plus or minus signs.
Later on we will be interested in writing the fermion operators as such powers of UODMs systematically, which will turn out to involve a certain non-uniqueness as can already be seen by means of the multiple ways one can choose the signs in this example (\ref{root_of_fermion}).
We denote by $D_L([x_1\,\,\cdots x_{N-1}]^T)$ an UODM with entries $x_1,\cdots,x_{N-1}$ on its possibly non-zero off-diagonal.
Here we view the components $x_{j}$ as an $N-1=2^L -1$ dimensional vector $[x_{j}]$ as this will be convenient later on, since if this vector is replaced by a linear combination of vectors, the corresponding $D_L$ will also be a linear combination.
In other words, we can view the space of the UODM as a vector space with standard basis $\tilde{E}_{j}$ and dimension $N-1=2s$, where $s$ is the spin of the representation that $\hat{S}_+$ belongs to.
In the following our main goal will be to construct a basis of this space whose elements $E^{(c)}_\alpha(L)$ will be build from the previously defined fermion operators as well as an alternative basis with elements $E^{(S)}_\alpha(s)$ constructed from spin operators.
After that we show how to express the spin operators in terms of $E^{(c)}_\alpha(L)$ and the fermion operators in terms of $E^{(S)}_\alpha(s)$.
Moreover, if we introduce a set of linearly independent UODMs
\begin{equation}
	E^{(b)}_j=\sum\limits_{k=1}^{N-1} (V_{(b)})_{jk} \tilde{E}_k
\end{equation}
with some $(N-1)\times (N-1)$ matrix $V_{(b)}$ defined by the general basis with label $b$, our $D_L$ can be written like
\begin{equation}\label{general_expansion}
	D_L([x_j])=\sum\limits_{j=1}^{N-1} x_j 
\tilde{E}_j=\sum\limits_{\substack{j=1\\k=1}}^{N-1} x_j 
(V^{-1}_{(b)})_{jk}E^{(b)}_k\,\,\,\text{.}
\end{equation}
The label $b$ here is a placeholder which will be replaced by either of the labels $c$ for the fermionic basis $\{E^{(c)}_j(L)\}_{\{j=1,\cdots,N-1\}}$ or $S$ for the spin-basis $\{E^{(S)}_j(s)\}_{\{j=1,\cdots,2s\}}$.
First we note, that we can write any UODM by means of the primitive two-dimensional fermion operators, and the projectors
\begin{equation}\label{Ps}
	\mathds{P}_\uparrow:=\hat{c}^\dagger \hat{c}=:\hat{n}\hspace{2cm} 
\mathds{P}_\downarrow := \hat{c}\hat{c}^\dagger\,\,\,\text{,}
\end{equation}
by starting at $L=1$ and then using, that we can place additional entries by combining the projectors with the other operators as
\begin{eqnarray}\label{L1_EXAMPLE}
	&D_1([x_1])=x_1 \hat{c}^\dagger
\end{eqnarray}
\begin{eqnarray}\label{L2_EXAMPLE}
	&D_2([x_1\,\,x_2\,\,x_3]^T)=
	\left[\begin{array}{cc}
		x_1\hat{c}^\dagger & x_2 \hat{c}\\
		\mathbf{0}_2 & x_3 \hat{c}^\dagger 
	\end{array}\right]=
\end{eqnarray}
\begin{eqnarray}\label{D2}
	&=\mathds{P}_\uparrow 
	\otimes D_1([x_1]) + x_2\hat{c}^\dagger\otimes \hat{c} + \mathds{P}_\downarrow \otimes D_1([x_3])\\
&D_1([x_1])\in\operatorname{Mat}(2^1,2^1)\hspace{1.5cm}D_2([x_1\,\,x_2\,\,x_3]
^T)\in\operatorname{Mat}(2^2,2^2)\nonumber
\end{eqnarray}
Here the first line (\ref{L1_EXAMPLE}) describes the case for UODMs of dimension $2^L \times 2^L |_{L=1}=2\times 2$, so with a single entry $x_1$, while below in (\ref{L2_EXAMPLE}) the analogous expression for UODMs of dimension $2^L \times 2^L |_{L=2}=4\times 4$, so with three entries $[x_1\,\, x_2\,\,x_3]$ is given.
Here and in the following we denote transposition of arrays by $T$.
The components $x_j$ of the respective expressions are just general names for the components of UODMs and are not literally related to each other, meaning that $x_1$ in (\ref{L1_EXAMPLE}) appears independently of $x_1$ in (\ref{L2_EXAMPLE}).
The latter expression (\ref{L2_EXAMPLE}) can be written in terms of the former as shown in (\ref{D2}), though there also arises a new term $x_2 \hat{c}^\dagger \otimes \hat{c}$.
This can be generalized by noting that every UODM can be seen as a matrix with blocks, that are UODMs themselves
\begin{eqnarray}\label{D_L_blocks}
	&D_L([x'_{j}]\oplus[x_{2^{L-1}}]\oplus[x''_{j}])=\left[\begin{array}
				{cc}
				D_{L-1}([x'_{j}]) & x_{2^{L-1}} \hat{c}^\dagger \otimes 
\hat{c}^{\otimes(L-1)}\\
				\mathbf{0}_{2^{L-1}\times 2^{L-1}} & D_{L-1}([x''_{j}])
												\end{array}\right]\,\,\,\text{.}\\
	&[x'_{j}],[x''_{j}] \in \mathbb{R}^{2^{L-1}-1}\nonumber
\end{eqnarray}
Writing this with the projection operators then gives the recursion relation
\begin{eqnarray}\label{D_L_recursion}
	&D_L([x_1 \,\, \cdots x_{2^L-1}]^T)=\mathds{P}_\uparrow \otimes 
D_{L-1}([x_1\,\,\cdots x_{2^{L-1}-1}]^T)\\
									&+x_{2^{L-1}} \hat{c}^\dagger \otimes 
\hat{c}^{\otimes(L-1)}
									+\mathds{P}_\downarrow \otimes 
D_{L-1}([x_{2^{L-1}+1}\,\, \cdots x_{2^L 
-1}]^T)\in\operatorname{Mat}(2^L,2^L)\nonumber
\end{eqnarray}
Since this is still written only in terms of the primitive operators $\hat{c}$, $\hat{c}^\dagger$ and their projectors we try to find a way of translating this into expressions depending on the $N$ dimensional operators $\hat{c}_\alpha$, $\hat{c}_\alpha^\dagger$. 
In order to illustrate, how this rewriting in terms of the higher dimensional operators can be done, we first state the general relations for the $2\times 2$ matrices
\begin{equation}\label{paulimult}
 \sigma_z \hat{c}=-\hat{c}\hspace{1cm} \hat{c}\sigma_z =\hat{c}
\end{equation}
\begin{equation}
  \sigma_z \hat{c}^\dagger=\hat{c}^\dagger \hspace{1cm} \hat{c}^\dagger\sigma_z 
=-\hat{c}^\dagger\hspace{1cm} \sigma_z^2=\mathds{1}_2
\end{equation}
and note, that the rewriting mathematically means the replacement of expressions involving tensor products by matrix products of higher dimensional operators.
In the following we want to give an explicit calculation for how one can rewrite a concrete tensor product expression built from the two-dimensional operators (\ref{CAR_1}) in terms of the previously defined higher dimensional fermionic operators (\ref{c_alpha}), (\ref{c_alpha_dagger}).
This calculation will also serve to motivate the specific form of the fermionic basis operators that we will arrive at later.
The concrete expression at the start of this calculation is just an arbitrarily picked one in order to showcase the types of calculations one has to undertake for deriving the formalism we present in this work.
Now consider for example for $L=3$ the following expression 
\begin{equation}
 \mathds{1}_2\otimes \hat{c}\otimes \mathds{1}_2
\end{equation}
It is not itself one of the fermionic operators $\hat{c}_\alpha$ defined above, but can be written with (\ref{CAR_1}) and (\ref{paulimult}) as
\begin{equation}\label{tech}
  \mathds{1}_2\otimes \hat{c}\otimes \mathds{1}_2=(\hat{c}\hat{c}^\dagger + 
\hat{c}^\dagger \hat{c})\otimes \hat{c} \otimes \mathds{1}_2=
\end{equation}
\begin{equation*}
  =\hat{c} \hat{c}^\dagger\otimes \hat{c} \otimes \mathds{1}_2+\hat{c}^\dagger 
\hat{c}\otimes \hat{c} \otimes \mathds{1}_2=
\end{equation*}
\begin{equation*}
  =(\hat{c}\otimes \mathds{1}_2 \otimes\mathds{1}_2) (\hat{c}^\dagger\otimes 
\mathds{1}_2 \otimes\mathds{1}_2)
  (-\sigma_z \otimes \hat{c} \otimes\mathds{1}_2)
  +
\end{equation*}
\begin{equation*}
  +(\hat{c}^\dagger \otimes \mathds{1}_2 \otimes\mathds{1}_2) (\hat{c}\otimes 
\mathds{1}_2 \otimes\mathds{1}_2)(\sigma_z \otimes \hat{c} \otimes\mathds{1}_2)=
\end{equation*}
\begin{equation}\label{example_calc_L3}
  =-\hat{c}_1 \hat{c}_1^\dagger \hat{c}_2 +\hat{c}_1^\dagger \hat{c}_1 
\hat{c}_2=(2\hat{c}_1^\dagger \hat{c}_1 -\mathds{1}_8)\hat{c}_2
\end{equation}
The operators $\hat{c}_\alpha$ and $\hat{c}_\alpha^\dagger$, $\alpha\in\{1,2\}$ in above equation (\ref{example_calc_L3}) are the operators (\ref{c_alpha}) and (\ref{c_alpha_dagger}) for $L=3$.
After having shown the techniques by which one can rewrite expressions involving tensor products of the operators (\ref{CAR_1}) in terms of the operators (\ref{c_alpha}), (\ref{c_alpha_dagger}), we want to go back to the expression for $D_2([x_1\,\,x_2\,\,x_3]^T)$ from equation (\ref{D2}) and apply the same types of computations.
Note, that hence in the following the operators $\hat{c}_\alpha$, $\hat{c}_\alpha^\dagger$ are the ones for $L=2$ as opposed to $L=3$ in (\ref{example_calc_L3}), so differring by a tensor product with $\mathds{1}_2$ from the right.
If we express all tensor-products in equation (\ref{L2_EXAMPLE}) by $\hat{c}_1$, $\hat{c}_2$ and their adjoints in the same fashion we obtain:
\begin{equation}\label{expansion_in_fermions1}
 x_1 \mathds{P}_\uparrow \otimes \hat{c}^\dagger + x_2 \hat{c}^\dagger \otimes 
\hat{c} + x_3 \mathds{P}_\downarrow \otimes \hat{c}^\dagger=
\end{equation}
\begin{equation*}
 =x_1 \hat{c}_1^\dagger \hat{c}_1 \hat{c}_2^\dagger - x_2 \hat{c}^\dagger_1 
\hat{c}_2 -x_3 \hat{c}_1\hat{c}_1^\dagger \hat{c}_2^\dagger=
\end{equation*}
\begin{equation}\label{example_calc}
 = (x_1+x_3) \hat{c}_1^\dagger \hat{c}_1 \hat{c}_2^\dagger - x_2 \hat{c}_1^\dagger 
\hat{c}_2 -x_3\hat{c}_2^\dagger
\end{equation}
In the last expression we can see linear combinations of the elements of the input vector $[x_j]$ summed over what we define as the fermionic basis operators serving as basis vectors of the vector space of $(2^L \times 2^L)|_{L=2}=4\times 4$ dimensional UODMs.
We now introduce the notation $E^{(c)}_\alpha(L)\in\operatorname{Mat}(2^L,2^L)$ for the $\alpha$-th fermionic basis vector build as a product of fermionic operators (hence the superscript $(c)$) with operators $\hat{c}_\alpha$ and $\hat{c}^\dagger_\alpha$ for $\alpha\in\{1,\cdots,L\}$, which can also be interpreted as that $E^{(c)}_\alpha(L)$ can be written as a $L$-fold tensor product of $2\times2$ operators, just like the $\hat{c}_\alpha$ and $\hat{c}^\dagger_\alpha$.
In the case of above example expression (\ref{example_calc}) for $L=2$ we obtain:
\begin{equation}\label{expansion_in_fermions2}
	D_2([x_1\,\,x_2 \,\,x_3]^T)=(x_1+ x_3)E^{(c)}_1(2) + (-x_2) E^{(c)}_2(2) + 
(-x_3) E^{(c)}_3(2)
\end{equation}
\begin{equation*}
	=\sum\limits_{\substack{j=1\\k=1}}^{N-1}x_{j} (V_{(\text{c})}^{-1}(2))_{jk} 
E^{(c)}_{k}(2)
\end{equation*}
\begin{equation}
	E^{(c)}_1(2):=\hat{c}_1^\dagger \hat{c}_1 \hat{c}_2^\dagger \hspace{2cm}
	E^{(c)}_2(2):=\hat{c}_1^\dagger \hat{c}_2\hspace{2cm}
	E^{(c)}_3(2):=\hat{c}_2^\dagger
\end{equation}
We choose not to include the signs arising in (\ref{expansion_in_fermions2}) in the basis elements but in the matrix $V_{(\text{c})}(2)$,which is the manifestation of the matrix $V_{(b)}$ from (\ref{general_expansion}) for the fermionic basis $(b)=(c)$.
We attached an additional argument $L$ to this matrix, in this case explicitly set to $L=2$, in order to denote the number of fermion flavors under consideration.
The basis transformation matrix from the fermionic basis $E^{(c)}_{j}(2)$ onto the standard one for this example is given by the self-inverse matrix
\begin{equation}\label{V_L2}
	V_{(\text{c})}^{T}(2)=\left[\begin{array}
			{ccc}
			1 & 0 & 1\\
			0 & -1 & 0\\
			0 & 0 & -1
			\end{array}\right]=(V_{(\text{c})}^{-1}(2))^T\,\,\,\text{.}
\end{equation}
For general $L$ this matrix $V_{(\text{c})}^T(2)$ will not be self-inverse anymore, so we continue to write down its inverse explicitly in all equations, since it is the inverse appearing in (\ref{general_expansion}) and not the matrix itself.
Here the additional transposition is introduced to adjust to the convention of treating component vectors as columns as again made clear later.
As one can see already in the recursion relation (\ref{D_L_recursion}), $D_L$ contains three essential parts for $L\geq2$: one part preceded by $\mathds{P}_\uparrow$ that takes care of the upper entries of the resulting UODM, the single middle entry obtained by $\hat{c}^\dagger\otimes \hat{c}^{\otimes (L-1)}$ and the part preceded by $\mathds{P}_\downarrow$ that takes care of the entries in the lower half of the matrix.
As shown explicitly for $L=2$ in (\ref{expansion_in_fermions1}), combined with (\ref{expansion_in_fermions2})  we can combine the recursion relation (\ref{D_L_recursion}) for $D_L(\cdots)$ with the general expansion (\ref{general_expansion}) in terms of the fermionic basis (\ref{expansion_in_fermions2}). 
We temporarily introduce the abbreviations $V^{-1}_{(c)}(L)=:P, V^{-1}_{(c)}(L-1)=:Q$ and $2^{L-1}=:K$, so that $2^L=:N=2K$.
As before in (\ref{D_L_blocks}) the component vector is split into three parts $[x_{j}^\prime]\oplus [x_K]\oplus[x_{j}^{\prime\prime}]$ of dimensions $K-1,1$ and $K-1$ respectively.
Furthermore we define the index-shift operation mapping from any of the indexed operators $\hat{a}_\alpha\in\{\hat{c}_\alpha,\hat{c}_\alpha^\dagger\}_{\alpha\in \{1,\cdots, L\}}$ (here the braces denote a set and not an anti-commutator) from $\operatorname{Mat}(2^L,2^L)$ to the corresponding operator from $\operatorname{Mat}(2^{L+1},2^{L+1})$:
\begin{eqnarray}\label{indshift}
	&\text{index-shift}(\hat{a}_{\alpha_1}\cdots \hat{a}_{\alpha_{g}}):=(\sigma_z\otimes \hat{a}_{\alpha_1})\cdots(\sigma_z \otimes \hat{a}_{\alpha_{g}})= \hat{a}_{\alpha_1+1}^\prime\cdots \hat{a}_{\alpha_{g}+1}^\prime,
\end{eqnarray}
which can be further rewritten as
\begin{eqnarray}
	&\hat{a}_{\alpha_1+1}^\prime\cdots \hat{a}_{\alpha_{g}+1}^\prime= \left\{  \begin{array}{r@{\quad}cr} 
	\sigma_z \otimes \hat{a}_{\alpha_1}\cdots \hat{a}_{\alpha_{g}} \hspace{1cm} \text{$g$ odd}\\
	\mathds{1}_2\otimes \hat{a}_{\alpha_1}\cdots \hat{a}_{\alpha_{g}} \hspace{1cm} \text{$g$ even}
	\end{array}\right. \\
	& = \sigma_z^{g}\otimes (\hat{a}_{\alpha_1}\cdots \hat{a}_{\alpha_{g}})\nonumber\\
&\hat{a}_\alpha\in\operatorname{Mat}(2^{L},2^{L})\hspace{1.5cm}\hat{a}_\alpha^\prime 
\in\operatorname{Mat}(2^{L+1},2^{L+1})\,\,\,\text{.}\nonumber
\end{eqnarray}
Please note, that we would denote powers of tensor products explicitly as $\sigma_z^{\otimes g}$, whereas the expression $\sigma_z^g$ is just intended to mean a $g$-fold matrix product of $\sigma_z$.
Then we can compute as follows:
\begin{eqnarray}\label{proof1_of_recursion}
	&D_L([x_{j}^\prime]\oplus [x_{K}] \oplus [x_{j}^{\prime\prime}])=
	\sum\limits_{\substack{j=1\\k=1}}^{K-1}x_{j}^\prime(P)_{jk}E^{(c)}_{k}(L) 
	+x_{K}(P)_{KK}E_{K}(L)+\\
&+\sum\limits_{\substack{j=1\\k=1}}^{K-1}x_{j}^{\prime\prime}(P)_{j+K\,\,\,k+K}E^{(c)}_
{k+K}(L)\nonumber\\
	&=\mathds{P}_\uparrow \otimes D_{L-1}([x_{j}^\prime])
	+x_{K} \hat{c}^\dagger \otimes \hat{c}^{\otimes(L-1)}+
	\mathds{P}_\downarrow \otimes D_{L-1}([x_{j}^{\prime\prime}])\nonumber
\end{eqnarray}
\begin{eqnarray}\label{proof_recursion_last_expression}
	&=\mathds{P}_\uparrow \otimes 
\sum\limits_{\substack{j=1\\k=1}}^{K-1}x_{j}^\prime(Q)_{jk}E^{(c)}_{k}(L-1)
	+(-1)^{L-1}x_{K} \hat{c}_1^\dagger\hat{c}_2 \cdots \hat{c}_L
\end{eqnarray}
\begin{eqnarray*}
	&+\mathds{P}_\downarrow 
\otimes\sum\limits_{\substack{j=1\\k=1}}^{K-1}x_{j}^{\prime\prime}(Q)_{jk}E^{(c)}_{k}(L-1)\\
\end{eqnarray*}
With (\ref{c_alpha}), (\ref{c_alpha_dagger}), (\ref{Ps}), (\ref{paulimult}) and (\ref{indshift}) we can deduce for $\hat{b}_j\in\operatorname{Mat}(2^L,2^L)$ and $\hat{n}_1$ here meaning the number operator (\ref{n_alpha}), though for dimension $2^{L+1} \times 2^{L+1}$
\begin{eqnarray}
	&\mathds{P}_\uparrow \otimes \hat{b}_j =c^\dagger c \otimes \hat{b}_j
	= (c^\dagger c \otimes \mathds{1}_{2^L})(\sigma_z^{f(b_j)}\otimes \hat{b}_j)=\hat{n}_1\operatorname{index-shift}(\hat{b}_j)
\end{eqnarray}
\begin{eqnarray}\label{b_j}
	&\mathds{P}_\downarrow \otimes \hat{b}_j =\hat{c}\hat{c}^\dagger \otimes \hat{b}_j= (-1)^{f(b_j)}(\hat{c}\hat{c}^\dagger \otimes \mathds{1}_{2^L})(\sigma_z^{f(b_j)}\otimes\hat{b}_j)=
\end{eqnarray}
\begin{eqnarray*}
	&=(-1)^{f(b_j)} (\mathds{1}_{2^{L+1}}-\hat{n}_1)\operatorname{index-shift}(\hat{b}_j)\,\,\,\text{,}
\end{eqnarray*}
where $f(\hat{b}_j)$ is the amount of matrix products of $\hat{c}_\alpha, \alpha\in\{1,\cdots,L\}$ and adjoints thereof (summarized as a general operator $\hat{a}_\alpha$ again) included in the general product operator $\hat{b}_j$.
For example for an operator $\hat{b}_j$ as in (\ref{b_j}):
\begin{eqnarray}
	&\hat{b}_j:=\hat{a}_1\cdots\hat{a}_{j},\,\,\, \hat{a}_\alpha\in \operatorname{Mat}(2^L,2^L),\,\,\,\alpha\in\{1,\cdots,L\}\\
	&f(b_j)=j\,\,\,\text{.}\nonumber
\end{eqnarray}
Remember, that the $\operatorname{index-shift}$ operation is used to rewrite a tensor product as a matrix product of higher-dimensional operators; the additional sign then arises here from $c^\dagger \sigma_z=-c^\dagger$.
In the following, $f_{k}$ always refers to the number of operators included in $E^{(c)}_{k}(L-1)$, $f_{k}:=f(E_{k}^{(c)}(L-1))$.
With this we obtain from the expression after the equality sign in (\ref{proof_recursion_last_expression})
\begin{eqnarray}\label{proof_recursion_very_last_expression}
&\sum\limits_{\substack{j=1\\k=1}}^{K-1}x_{j}^\prime(Q)_{jk}\hat{n}_1\text{index-shift}
\left(E^{(c)}_{k}(L-1)\right)
	+(-1)^{L-1}x_{K} E^{(c)}_K(L)\\
&+\sum\limits_{\substack{j=1\\k=1}}^{K-1}x_{j}^{\prime\prime}(Q)_{jk}(-1)^{f_{k}}
(\mathds{1}_{N} - \hat{n}_1)\text{index-shift}\left(E^{(c)}_{k}(L-1)\right)\nonumber\\
&=\sum\limits_{\substack{j=1\\k=1}}^{K-1}(x_{j}^\prime+(-1)^{f_{k}+1}x_{j}^{\prime\prime
})(Q)_{jk}\hat{n}_1\text{index-shift}\left(E^{(c)}_{k}(L-1)\right)+\nonumber\\
	&+(-1)^{L-1}x_{K} E^{(c)}_K(L)
+\sum\limits_{\substack{j=1\\k=1}}^{K-1}(-1)^{f_{k}}x_{j}^{\prime\prime}(Q)_{jk}\text
{index-shift}\left(E^{(c)}_{k}(L-1)\right)\nonumber
\end{eqnarray}
Furthermore from the knowledge that the first term $D_{L-1}([x_{j}^\prime])$ in the second line creates the upper block of the resultant $D_L(\cdots)$ and this first term is exclusively written with basis elements $\hat{n}_1\text{index-shift}\left(E^{(c)}_{k}(L-1)\right)$, we can conclude that these have to be the new basis elements $E^{(c)}_{k}(L)$.
Analogously we find the recursive expressions 
\begin{equation}\label{recursion_relations}
E^{(c)}_\alpha(L)=\hat{n}_1\text{index-shift}(E^{(c)}_\alpha(L-1))\,\,\,\,
\alpha\in\left\{1,\cdots,K-1\right\}
\end{equation}
\begin{equation}\label{recursion_relations2}
	E^{(c)}_{K}(L)=\hat{c}^\dagger_1\hat{c}_2\cdots \hat{c}_{L}
\end{equation}
\begin{equation}\label{recursion_relations3}
E^{(c)}_{\alpha+K}(L)=\text{index-shift}(E^{(c)}_\alpha(L-1))\,\,\,\,\alpha\in\left\{1,\cdots,K-1\right\}
\end{equation}
for the general basis operators for the space of UODMs, that are made of fermionic operators.
In the appendix (\ref{A.2.}) we prove the linear independence of the so defined basis operators and illustrate the application of the recursion relations (\ref{recursion_relations})-(\ref{recursion_relations3}). 
By comparing the first (\ref{proof1_of_recursion}) and last (\ref{proof_recursion_very_last_expression}) lines of the previous computation we can determine the matrix $V_{(c)}^{-1}(L)$ in terms of $V_{(c)}^{-1}(L-1)$ which can be written with $K:=2^{L-1}$ again, in block form as
\begin{eqnarray}
	V^{-1}_{(\text{c})}(L)=
	\left[
	\begin{array}
	{ccc}
	V_{(c)}^{-1}(L-1) & \mathbf{0}_{(K-1)\times 1} & \mathbf{0}_{K-1\,\,K-1}\\
	\mathbf{0}_{1\times (K-1)} & (-1)^{L-1} & \mathbf{0}_{1\times (K-1)}\\
	 A & \mathbf{0}_{(K-1)\times 1} & B 
	\end{array}
	\right]\\
	A:=\left[(-1)^{f_{k}+1} \left(V_{(c)}^{-1}(L-1)\right)_{jk}\right]\\
	B:=\left[(-1)^{f_{k}} 
\left(V_{(c)}^{-1}(L-1)\right)_{jk}\right]\,\,\,\text{.}
\end{eqnarray}
Note, that the expansion $D_L(\cdots)$ is written in matrix form with an additional transposition as
\begin{eqnarray}\label{transpose_operation}
	&D_L([x_{j}])=\sum\limits_{\substack{j=1\\k=1}}^{N-1} x_{j} 
(V_{(c)}^{-1})(L)_{jk}E^{(c)}_{k}(L)=\\
	&=\left[ E^{(c)}_1(L) \cdots E^{(c)}_{N-1}(L)\right] (V_{(c)}^{-1})^T(L) 
	\left[
	\begin{array}
	{c}
	x_1\\
	\vdots\\
	x_{N-1}
	\end{array}
	\right] \,\,\,\text{,}\nonumber
\end{eqnarray}
since we follow the convention that component vectors $[x_{j}]$ are written as columns as opposed to rows.
We are now in a position to expand the spin ladder operator in terms of fermionic operators by simply replacing the component vector in $D_L([x_{j}])$ by the non-zero entries of the spin operator as alluded to above.
For brevity we give only the results for the three dimensional expansion $L=2$ and refer to the appendix (\ref{A.4.}) for $L=3$ and $L=4$.\\
\\

Spin $3/2$ corresponding to $L=2$ and with $x_{j}(\hat{S}_+)$ denoting the needed upper off-diagonal components of the matrix $\hat{S}_+$:
\begin{eqnarray}\label{spin32exp}
	&[x_{j}(\hat{S}_+)]=[(\hat{S}_+)_{j,j+1}]=\left[\sqrt{3}\,\,\,2\,\,\,\sqrt{3}\right]^T
\end{eqnarray}
\begin{eqnarray}\label{62} 
	&\hat{S}_+=D_2([x_{j}(\hat{S}_+)])=
\end{eqnarray}
\begin{eqnarray*}
	&=\left[E^{(c)}_1(2)\,\,\, E^{(c)}_2(2)\,\,\, E^{(c)}_3(2)\right] \left[\begin{array}
			{ccc}
			1 & 0 & 1\\
			0 & -1 & 0\\
			0 & 0 & -1
	\end{array}\right]
	\left[\begin{array}{c}
		\sqrt{3}\\
		2\\
		\sqrt{3}
	\end{array}\right]=
\end{eqnarray*}
\begin{eqnarray*}
	&=2\sqrt{3}\hat{n}_1 \hat{c}_2^\dagger -2\hat{c}_1^\dagger \hat{c}_2 -\sqrt{3} \hat{c}_2^\dagger\\
	&\hat{S}_+\in\operatorname{Mat}(4,4)\hspace{1.5cm} \hat{c}_\alpha\in\operatorname{Mat}(4,4)
\end{eqnarray*}
Here the $3\times 3$ matrix in (\ref{62}) is given by $(V_{(c)}^{-1})^T(2)$ and the fermionic basis operators $E^{(c)}_\alpha(2)$, which are themselves $4\times 4$ matrices, are computed according to (\ref{recursion_relations}), (\ref{recursion_relations2}), (\ref{recursion_relations3}).
There was a slightly different but equivalent equation obtained in \cite{Dobrov}, though without the general formalism applicable for arbitrary numbers $L$.
Also it should be noted, that there are multiple possibilities to define matrices that fulfill the $2^L$-dimensional CAR (\ref{CAR}), differing by permutations of tensor-product factors.

\subsection{Rewriting diagonal Spin Operators}

We state, that for diagonal operators like the spin-z operator we even get a closed form expression in terms of the fermionic operators by first computing the expansion of $\hat{S}_+$ in terms of fermionic operators and then using (\ref{su2}) to find the corresponding expansion of $\hat{S}_z$.
By inspecting the structure of the results for $L=2$ and $L=3$ we could first assume a general formula that contains them as special cases and afterwards show that it indeed holds.
In (\ref{Szexplicit}) below we present the so obtained expansion of the spin-z operator for any spin $s$ fulfilling (\ref{compatibility}) in terms of the corresponding number $L(s)=\log_2(2s+1)$ of fermion operators, which are used here in order to obtain the number operators as $\hat{n}_\alpha=\hat{c}^\dagger_\alpha \hat{c}_\alpha$.\\
\\
\textbf{Proposition:}
For the spin-z operator $\hat{S}_z$ in a spin $s$ representation we have the expansion in terms of the $(2s+1)\times (2s+1)=2^L \times 2^L$ dimensional fermion number operators (\ref{n_alpha}):
\begin{eqnarray}\label{Szexplicit}
\hat{S}_z=-s\mathds{1}_{2^{L(s)}}+\sum\limits_{j=1}^{L(s)}2^{j-1}\hat{n}_{
L+1-j},\,\,\, L(s):=\log_2(2s+1)\\
	\hat{S}_z,\,\,\hat{n}_\alpha\in\operatorname{Mat}(2s+1,2s+1)\nonumber
\end{eqnarray}
\\
\textbf{Proof}:\\
In order to check this equation (\ref{Szexplicit}) we rewrite it in the form
\begin{eqnarray}\label{proof1}
\hat{S}_z+s\mathds{1}_{2^{L(s)}}=\sum\limits_{j=1}^{L(s)=\log_2(2s+1)}2^{j-1}\hat{n}_{
L+1-j}
\end{eqnarray}
and note that
\begin{eqnarray}
	\hat{S}_z + s\mathds{1}_{2^{L(s)}}=\operatorname{diag}((2s,2s-1,\cdots, 
0))\,\,\,\text{.}
\end{eqnarray}
The right hand side of (\ref{proof1}) can be noticed to be just the binary representation of the decreasing sequence of diagonal entries $2s, 2s-1,\cdots$.
For example, for the $L(3/2)=2$ case we have
\begin{eqnarray}
	\sum\limits_{j=1}^{L(\frac{3}{2})=2}2^{j-1}\hat{n}_{L+1-j}=\hat{n}_{2} +2 \hat{n}_{1}=\\
	=\operatorname{diag}\left(\left[2^1\,\,\,2^0\right]
	\left[\begin{array}{cccc}
		(\hat{n}_1)_{11} & (\hat{n}_1)_{22} &(\hat{n}_1)_{33} & (\hat{n}_1)_{44}\\
		(\hat{n}_2)_{11} & (\hat{n}_2)_{22}& (\hat{n}_2)_{33} & (\hat{n}_2)_{44}
	\end{array}\right]\right)=\nonumber\\
	=\operatorname{diag}\left(\left[2^1\,\,2^0\right]
	\left[\begin{array}{cccc}
		1 & 1 & 0 & 0\\
		1 & 0 & 1& 0
	\end{array}\right]\right)=\nonumber\\
	=\operatorname{diag}\left(\left[ 3 \,\,\, 2 \,\,\, 1 \,\,\, 0\right]\right)=
	\operatorname{diag}\left(\left( 3 , 2 , 1 , 0\right)\right)\nonumber
\end{eqnarray}
Here the notations with $\operatorname{diag}$ applied to a row- or column-vector as well as applied to a tuple are understood to mean the same.
The structure of the number operators is given in terms of $2^\alpha$ blocks of respectively size $2^{L-\alpha}$, alternating between 0 and 1, so that they yield $2^\alpha 2^{L-\alpha}=2^L$ diagonal components,
\begin{eqnarray}
\hat{n}_\alpha=\operatorname{diag}((\underbrace{1,\cdots,1}_{2^{L-\alpha}},
\underbrace{0,\cdots,0}_{2^{L-\alpha}},\underbrace{1,\cdots,1}_{2^{L-\alpha}},
\underbrace{0,\cdots,0}_{2^{L-\alpha}},\cdots))\,\,\,\text{.}
\end{eqnarray}
We hence obtain all binary representations of numbers starting from the maximum number
\begin{eqnarray}
\sum\limits_{j=0}^{L-1}2^{j}=\frac{1-2^{L-1+1}}{1-2}=\frac{1-2^{\log_2(2s+1)}}{-1}=2s
\end{eqnarray}
to 0 for the general $L$ fermion case.
\newline
$\null\nobreak\hfill\square$
\newline

\subsection{Rewriting diagonal Fermion Operators}
In order to obtain the reverse exact representation of the number operators in terms of powers $\hat{S}_z^k$ we note that for spin $s$ we have $2s+1$ linearly independent powers, starting at 0, because the minimal polynomial \cite{MINIMAL_POLYNOMIAL} of $\hat{S}_z$ in the spin $s$ representation is of degree $2s+1$.
Therefore for diagonal operators of dimension $2s+1$, like the number operators $\hat{n}_\alpha\in\operatorname{Mat}(2^L,2^L)$ with $2^L=2s+1$ we can use $\{\hat{S}_z^k\}_{k\in\{0,1\cdots,2s\}}$ as a basis to expand them with real coefficients $\tilde{y}^{(\alpha)}_\beta$:
\begin{eqnarray}\label{diagonal_expansion}
	&\hat{n}_\alpha=\sum\limits_{\beta=1}^{2s+1} 
\tilde{y}^{(\alpha)}_{\beta}\hat{S}_z^{\beta-1}
\end{eqnarray}
\begin{eqnarray}\label{y_beta}
&[\tilde{y}^{(\alpha)}_\beta]:=\mathcal{A}^{-1}[(\hat{n}_\alpha)_{\beta\beta}]
\in\mathbb{R}^{2s+1}\\
	&\mathcal{A}:=\left[\begin{array}
					{cccc}
					(\hat{S}_z^0)_{11} & (\hat{S}_z^1)_{11} & \cdots & 
(\hat{S}_z^{2s})_{11}\\
					(\hat{S}_z^0)_{22} & (\hat{S}_z^1)_{22} & \cdots & 
(\hat{S}_z^{2s})_{22}\\
					\vdots & \vdots & \vdots &\vdots \\
					(\hat{S}_z^0)_{2s+1\,\,2s+1} & (\hat{S}_z^1)_{2s+1\,\,2s+1} & 
\cdots & (\hat{S}_z^{2s})_{2s+1\,\, 2s+1}
				\end{array}\right]\,\,\,\text{,}\nonumber\\
&\mathcal{A}\in\operatorname{Mat}(2s+1,2s+1)\,\,\,\text{.}\nonumber
\end{eqnarray}
Here the matrix $\mathcal{A}$ can be recognized as the Vandermonde Matrix $\operatorname{Vand}([(\hat{S}_z)_{kk}])$ of the vector of diagonal components of $\hat{S}_z$.
For these matrices there is an analytic expression for their inverse 
\cite{VANDERMONDE_INVERSE}, \cite{VANDERMONDE_INVERSE2_AND_DEF}:
\begin{eqnarray}\label{vand_inverse}
(\operatorname{Vand}^{-1}([v_{j}]))_{jk}=(-1)^{j+k}\frac{\mathcal{S}_{2s+1-j\,\,,k}}{
\prod\limits_{m<n}^{2s+1} (v_{n} - v_{m})}\,\,\, j,k=1,\cdots 2s+1\,\,\,\text{,}
\end{eqnarray}
where $\mathcal{S}_k$ are the Elementary Symmetric Polynomials well known from  the mathematical literature \cite{ESP}(note, that here we use the slightly different notation from \cite{VANDERMONDE_INVERSE} however)
\begin{eqnarray}
	& \mathcal{S}_k:=\mathcal{S}_k(v_1,\cdots,v_k) =\sum\limits_{1\leq j_1<\cdots<j_k\leq N}^{N} v_{j_1}v_{j_2}\cdots v_{j_k}\hspace{1cm} 1\leq k \leq N:=2^L\nonumber\\
	\hfill\\
	&\mathcal{S}_0(v_1,\cdots,v_k):=1\nonumber\\
	&\mathcal{S}_{k,j}:=\mathcal{S}_k(v_1,\cdots 
v_{j-1},v_{j+1},\cdots,v_N)\,\,\,\text{.}\nonumber
\end{eqnarray}
Replacing the general variables $v_i$ by the diagonal components $[(\hat{S}_z)_{jj}]^T=[s\,\,s-1\cdots\,\,-s]^T$ we can write the expansion of the fermionic number operator by the diagonal spin operators as 
\begin{eqnarray}\label{n_Sz_reversed}
&\hat{n}_\alpha=p_\alpha^{(L(s))}(\hat{S}_z):=\sum\limits_{\beta=1}^{2s+1}\left(\mathcal{A}^{-1}[
(\hat{n}_\alpha)_{jj}]\right)_\beta \hat{S}_z^{\beta-1}
\end{eqnarray}
\begin{eqnarray}\label{n_Sz_reversed_end}
&(\mathcal{A}^{-1})_{jk}=(-1)^{j+k}\frac{\mathcal{S}_{2s+1-j\,\,,k}(s,\cdots,-s)}{
\prod\limits_{\substack{m<n\\ m=k \text{ or } n=k}}^{2s+1} (m-n)}=
\end{eqnarray}
\begin{eqnarray}
&=(-1)^{j+k+s(2s+1)}\frac{\mathcal{S}_{2s+1-j\,\,,k}(s,\cdots,-s)}{(2s+1-k)!k!},\,\,\,\,j,k=1,\cdots,2s+1\nonumber
\end{eqnarray}
where the denominator in the product is obtained by $(m-n)=(s-n+1)-(s-m+1)=(\hat{S}_z)_{nn}-(\hat{S}_z)_{mm}$ and $\prod\limits_{\substack{m<n\\ m=k \text{ or } m=n}}^{2s+1} (m-n)=(2s+1-k)!k!$. 
By using these equations (\ref{n_Sz_reversed})-(\ref{n_Sz_reversed_end}) we can express all number operators $\hat{n}_\alpha(L)$, here written with an explicit specification of the given $L$, by above (\ref{n_Sz_reversed}) polynomials $\hat{n}_\alpha(L)=p^{(L(s))}_\alpha(\hat{S}_z)$ of order $2^L=2s+1$ in $\hat{S}_z$ and again with $L(s):=\log_2(2s+1)$.

\subsection{Mapping Fermions onto Spins}
In this section we want to present the mapping in the reverse direction, i.e. from the operators $\hat{S}_\pm$, $\hat{S}_z$ onto the fermionic operators $\hat{c}_\alpha$, $\hat{c}_\alpha^\dagger$.
Similarly to before we only need to express the creation operator, since the respective annihilation operator follows by means of taking adjoints, but now we have a number of $L$ many of those, $\hat{c}_1^\dagger,\cdots,\hat{c}_L^\dagger$, as opposed to the single spin creation operator $\hat{S}_+$.
Additionally we face the difficulty, that ,for a given $L$, always only one of the fermion operators $\hat{c}_\alpha^\dagger$ is an UDOM itself, namely the last one, $\hat{c}_L^\dagger$. 
This problem can be alleviated by the already stated fact, that nevertheless they all can be written as powers $M_\alpha^{2^{L-\alpha}}$ of an UODM with off-diagonal components $(\hat{c}_\alpha^\dagger)^{\frac{1}{2^{L-\alpha}}}$.
Since these matrix roots are non-unique, there are multiple possible components for a given $\hat{c}_\alpha^\dagger$ as again already shown in the explicit example (\ref{root_of_fermion}).
We now proceed in the same fashion as in the previous section by finding a basis built from the spin operators so that we can write the general UODM $D_L(\cdots)$ in terms of it.
To achieve this we only need to make use of the fact, that in a spin $s$ representation we have at most $2s+1$ linearly independent operator powers $\hat{S}_z$ which are diagonal. 
Multiplying with $\hat{S}_+$ we can turn the complete set of these operator powers into the right amount of $N-1$ basis elements to expand a UODM in.
Hence the spin operator basis to replace the $E^{(c)}_\alpha(L)$ is given by
\begin{equation}\label{spin_basis}
	E^{(S)}_\alpha(s):=\hat{S}_+\hat{S}_z^{\alpha-1}\in 
\operatorname{Mat}(2s+1,2s+1)\,\,\,\text{.}
	\end{equation}
An advantage over the reverse mapping is the ability to express the analogue of the matrix $V_{(c)}$, which we call $V_{(S)}$, non-recursively, since we know that the components of the above stated basis (\ref{spin_basis}) are\cite{CLEBSCH_GORDAN}:
\begin{eqnarray}
	(E^{(S)}_\alpha(s))_{jk}=\sqrt{(s+j)(s-j+1)}(j)^{\alpha-1} 
\delta_{j\,\,k-1}\text{,}\,\,\, j,k\in \{s,s-1,\cdots,-s\}\,\,\,\text{.}\nonumber\\
\end{eqnarray}
This then gives explicitly
\begin{equation}\label{def_Vs}
	(V_{(S)})_{jk}(s)=(E^{(S)}_{k}(s))_{j\,\,j+1}\in\operatorname{Mat}(2s,2s)\,\,\,\text{.}
\end{equation}
with the defining equation for $V_{(S)}$ given by
\begin{equation}\label{uodm_expansion}
	M_\alpha=D_L([x^{(c,\alpha)}_{j}])=\sum\limits_{j=1}^{N-1} 
x^{(c,\alpha)}_{j}\tilde{E}_{j}=\sum\limits_{\substack{j=1\\k=1}}^{N-1} x^{(c,\alpha)}_{j} (V_{(S)}^{-1})_{jk}(s)
E^{(S)}_{k}(s)\,\,\,\text{.}
\end{equation}
While we now have an analytical expression for the basis transformation matrix as well as for all basis vectors $E^{(S)}_\alpha(s)$ the same does not hold for the coefficient vectors $[x^{(c,\alpha)}_{j}]$, carrying an additional superscript $\alpha$ to reference the respective fermion operator, like before for $[x^{(S)}_{j}]$.
In order to compute the components of the fermionic operators one needs to find a UODM solution $M_\alpha$ of
\begin{equation}\label{fermion_expansion}
	(\hat{c}_\alpha^\dagger)^{\frac{1}{2^{L-\alpha}}}=M_\alpha\,\,\,\text{.}
\end{equation}
By using the fact, that taking products of UODMs shifts their non-zero entries one space to the right, combined with the knowledge that the entries of the fermionic operator $\hat{c}_\alpha^\dagger$ are positioned in the $L-\alpha+1$-st off-diagonal we know, that we can reach the corresponding off-diagonal of the matrix $\hat{c}_\alpha^\dagger$ by taking $2^{L-\alpha}$ products of UODMs with entries only on the first off-diagonal, which results in taking the just stated power $2^{L-\alpha}$ to go in the reverse direction.
Had we defined the basis vectors $E^{(S)}_\alpha(s)$ with a variable power of $\hat{S}_+$, it would not be necessary to compute these matrix roots, but the price for this is that then the vector spaces under consideration would also no longer just be given by the UODM, resulting in expansions in different dimensions for the two directions of our mapping.
Formally this would then correspond to a mapping between (subspaces of) the universal enveloping algebras of the algebras under study so far.
One can deduce the signs in the non-zero off-diagonal components with the adjunct map
\begin{equation}
	\lambda:a\mapsto a \oplus -a = (a,-a)\,\,\,\text{.}
\end{equation}
One can write the tensor products of Pauli matrices as
\begin{equation}
	\sigma_z^{\otimes (\alpha-1)} = \operatorname{diag}(\lambda^{\circ 
(\alpha-1)}(1))\,\,\,\text{,}
\end{equation}
where $\circ$ denotes function composition.
With the tensor product $\mathds{1}_2^{\otimes 
(L-\alpha)}=\mathds{1}_{2^{L-\alpha}}$ we then can write
\begin{equation}
	\hat{c}_\alpha^\dagger=\operatorname{diag}(\lambda^{\circ (\alpha-1)}(1))\otimes 
\hat{c}^\dagger \otimes \mathds{1}_{2^{L-\alpha}}\,\,\,\text{.}
\end{equation}
The first part gives all the signs to appear in $M$ while the number of zeros in between the signed parts increases with the amount of identity matrices in the tensor product.
For example, we get for the expansion coefficients for $L=2$ and $L=3$ by computing the sequences of $1$ and $-1$ with this map $\lambda$ we obtain the vectors of expansion coefficients $[x^{(c,\alpha)}_{j}]$ of the UODM $M_\alpha$ from (\ref{fermion_expansion}) as
\begin{eqnarray}\label{ferm_comp}
	&\text{$L=2$}\\
	& \lambda^{\circ 0}(1)=(1) & [x_{j}^{(c,1)}]=[1\,\,\,1\,\,\,1]^T\nonumber \\
	& \lambda^{\circ 1}(1)=(1,-1) & [x_{j}^{(c,2)}]=[1\,\,\,0\,\,\,-1]^T \nonumber\\
	&\text{$L=3$}\\
	&\lambda^{\circ 0}(1)=(1) & 
[x_{j}^{(c,1)}]=[1\,\,\,1\,\,\,1\,\,\,1\,\,\,1\,\,\,1\,\,\,1]^T\nonumber\\
	&\lambda^{\circ 1}(1)=(1,-1) & 
[x_{j}^{(c,2)}]=[1\,\,\,1\,\,\,1\,\,\,0\,\,\,-1\,\,\,1\,\,\,-1]^T\nonumber
\end{eqnarray}
\begin{eqnarray*}\label{ferm_comp_end}
	&\lambda^{\circ 2}(1)=(1,-1,-1,1) & 
[x_{j}^{(c,3)}]=[1\,\,\,0\,\,\,-1\,\,\,0\,\,\,-1\,\,\,0\,\,\,1]^T\,\,\,\text{.}
\end{eqnarray*}
From this one obtains the above written component vectors $[x_{j}^{(c,\alpha)}]$ as follows: for every number in the tuple obtained by application of $\lambda$ we have another section of $1$ and $-1$ in the component vectors on the right that together have to multiply pairwise to the number inside the result of $\lambda$, simulating the products of elements of UODMs.
Every such section in the $[x_{j}^{(c,\alpha)}]$ is separated from the following section by a number $0$.
We can illustrate the meaning of these expansion coefficients by looking at, e.g., $[x_{j}^{(c,2)}]$ and writing down the list of numbers obtained from pairwise multiplication of each neighboring numbers in $[x_{j}^{(c,2)}]$:
\begin{eqnarray}
[1\,\,\,\,1\,\,\,1\,\,\,0\,\,\,-1\,\,\,1\,\,\,-1]^T\rightarrow[1\,\,\,\,1\,\,\,0\,\,\,
0\,\,\,-1\,\,\,-1]^T\,\,\,\text{.}
\end{eqnarray}
The second list of numbers are the entries of the first off-diagonal of the operator $\hat{c}_2$, which contains its only non-zero entries.By means of these simple rules one can in principle determine all expansion coefficients for arbitrary operator dimensions, but since, apart from the respective last vector $[x^{(c,L)}_{j}]$, the coefficient vectors $[x_{j}^{(c,\alpha)}]$ are the entries of a matrix root, they are only determined up to a sign for each respective block in the component vectors.
This sign is not fixed explicitly by our formalism, just as taking the root of a number does not fix the corresponding sign of the result either, but since the concrete choice of allowable signs is irrelevant for the purpose of the formalism, we do not see this as much of a downside in itself.
On the other hand, a multi-valuedness as just mentioned implies, that it does not seem feasible to give an analytical closed form expression for the component vectors and instead we content ourselves with above recursive definition.
We are now prepared to give the reverse expressions to (\ref{62}) by making use of (\ref{uodm_expansion}) and (\ref{fermion_expansion}).
Note that in the explicit examples (\ref{c1dagger_expansion}), (\ref{c2dagger_expansion}) below, the translation of (\ref{uodm_expansion}) into an index-free equation of arrays leads to the transpose of the $3\times 3$ matrix $V_{(S)}^{-1}(s=3/2)$ in the middle of the equations.
Here we only give the explicit expressions for $L=2$ including the spin basis operators (\ref{spin_basis}) for $s=3/2$:
\begin{eqnarray}\label{c1dagger_expansion}
	& 
\hat{c}_1^\dagger=\left(\left[E^{(S)}_1\left(\frac{3}{2}\right)\,\,\,E^{(S)}
_2\left(\frac{3}{2}\right)\,\,\,E^{(S)}_3\left(\frac{3}{2}\right)\right]
	\left[\begin{array}{ccc}
	\frac{3}{8\sqrt{3}} & \frac{3}{8} & \frac{-1}{8\sqrt{3}}\\
	\frac{1}{\sqrt{3}} & \frac{-1}{2} & 0\\
	\frac {1}{2\sqrt{3}} & \frac{-1}{2} & \frac{1}{2\sqrt{3}}
\end{array}\right]
	\left[\begin{array}{c}
			1\\
			1\\
			1
\end{array}\right]\right)^2\\
	&=\left(\frac{2+3\sqrt{3}}{8\sqrt{3}}\hat{S}_+ 
+\frac{2-\sqrt{3}}{2\sqrt{3}}(\hat{S}_+ \hat{S}_z + \hat{S}_+ 
\hat{S}_z^2)\right)^2\nonumber
\end{eqnarray}
\begin{eqnarray}\label{c2dagger_expansion}
& \hat{c}_2^\dagger=\left[E^{(S)}_1\left(\frac{3}{2}\right)\,\,\,E^{(S)}
_2\left(\frac{3}{2}\right)\,\,\,E^{(S)}_3\left(\frac{3}{2}\right)\right]
	\left[\begin{array}{ccc}
			\frac{3}{8\sqrt{3}} & \frac{3}{8} & \frac{-1}{8\sqrt{3}}\\
			\frac{1}{\sqrt{3}} & \frac{-1}{2} & 0\\
			\frac {1}{2\sqrt{3}} & \frac{-1}{2} & \frac{1}{2\sqrt{3}}
	\end{array}\right]
	\left[\begin{array}{c}
	1\\
	0\\
	-1
	\end{array}\right]\\
	&=\frac{4}{8\sqrt{3}} \hat{S}_+ + \frac{1}{\sqrt{3}} \hat{S}_+ \hat{S}_z\,\,\,\text{.}\nonumber
\end{eqnarray}
The expressions for $L=3$ are computed analogously and can be found in the appendix (\ref{A.5.}).

\subsection{Solution of the Recursion Relation}
In this section we give and motivate our solution for the recursion relation for 
$D_L(\cdots)$.
It reads:
\begin{eqnarray}\label{solrecu}
&D_L([x_{j}])=\sum\limits_{j=1}^{N-1}x_{j}\left(\bigotimes\limits_{l=1}^{N-1}\mathcal{O}^
{\left((\mathcal{P}_{L,l})_{j}\right)}\right)\otimes\hat{c}^\dagger\otimes 
\hat{c}^{\otimes m_{j}}\text{,}\\
&\Pi_{j}:=\bigotimes\limits_{l=1}^{N-1}\mathcal{O}^{\left((\mathcal{P}_{L,l})_{j}\right)}
\\
&\mathcal{O}^{(1)}:=\mathds{P}_\uparrow\text{,}\hspace{1.5cm} 
\mathcal{O}^{(0)}:=1\text{,}\hspace{1.5cm} \mathcal{O}^{(-1)}:=\mathds{P}_\downarrow 
\\
&\mathcal{P}_{L,l}:=\bigoplus\limits_{k=1}^{l-1}(r_{L-l}\oplus 0) \oplus r_{L-l} 
\\
&r_L:=[\underbrace{1\,\,\,1\,\,\,\cdots\,\,\,1\,\,\,1}_{L 
-1}\,\,\,0\,\,\,\underbrace{-1\,\,\,-1\,\,\,\cdots\,\,\,-1\,\,\,-1}_{L 
-1}]^T\\
&m_{j}:=L-1-\log_2(\operatorname{dim}_1(\Pi_{j}))\text{,}\,\,\, 
\operatorname{dim}_1(M\in\operatorname{Mat}(q,p))=q
\end{eqnarray} 
Here the operators $\mathcal{O}^{(\alpha)}$ are used to express general arrangements $(\uparrow,\cdots \uparrow,\downarrow,\cdots,\downarrow,\uparrow,\cdots)$ of combinations of $\mathds{P}_{\uparrow}$,$\mathds{P}_{\downarrow}$ which appear in all possible ordered combinations and for lengths of arrangements between $N-1$ and $0$ in $D_L$.
The difficulty is not to write down all general summands of $D_L$, but to assign them to a single index in an organized manner, so that the right summands can be summed with the components $x_{j}$ of the coefficient vector.
For this purpose we devise a pattern matrix $\mathcal{P}_L$ that is of dimension $(2^L -1) \times (L-1)$ and whose $k$-th row $(\mathcal{P}_{L,l})_k$ contains all the labels $\alpha$ to determine, which tensor product combination of $\mathcal{O}^{(\alpha)}$ operators is to be used in the respective summand that is combined with $x_l$.
When $\alpha=0$ is inserted, these operators are defined to be just scalars $\mathcal{O}^{(0)}:=1\in\mathbb{R}$ so that the complete tensor product is not affected by them.
This is necessary for the summands in $D_L$ in which a lower amount of $\mathds{P}_{\uparrow /\downarrow}$ operators appears.
Since the amount of $2\times 2$ operators in the tensor product has to stay at $N$ for all terms, these summands are filled up with additional $\hat{c}$ operators of the right amount $m_{j}$.
The value of $m_{j}$ is determined by the lack in dimension of the previous tensor-product $\Pi$, which is of dimension $2^{L-1-(\text{number of ($\alpha=0$) terms})}$.
The operator $\hat{c}^\dagger$ always appears in every summand in the middle of the two parts $\Pi$ and $\hat{c}^{\otimes m}$, so is inserted statically. 
We give explicit examples of the application of (\ref{solrecu}) in the appendix (\ref{A.6.}).

\section{Applications}
\subsection{Diagonalization of Spin- and Fermionic Hamiltonians}
Note the difference of our mapping in comparison to other known spin-fermion replacements, where the representation is usually restricted to $s=1/2$ like in the Jordan-Wigner transformation \cite{JWT} or the mapping is expressed in terms of constrained fermionic operators that fulfill a slightly different but equivalent form of CAR \cite{General_JWT_Batista}.
Also while in those other approaches based on a whole chain of spins, it is possible to reduce the chain length to one as a special case, our formalism is fundamentally based on describing just a single large spin in terms of fermions or vice versa.
In this sense our formalism might be seen as a form of coupling theory for fermion algebras, similarly to how couplings of spin algebras yields the coupling theory from standard $\mathfrak{su}(2)$ representation theory including the Clebsch-Gordan coefficients\cite{CLEBSCH_GORDAN}.
The difference to the topic of spin-coupling theory from ordinary quantum mechanics is that our system is not composed of smaller spins that yield a direct sum of larger spins, but of objects belonging to an other algebra, namely the fermion algebra.
Inside of this representation space we have to identify the spin- and occupation number bases as for example for $s=3/2$:
\begin{eqnarray}
	|\frac{3}{2},\frac{3}{2}\rangle \mapsto|1,1\rangle\,\,\,\,
	|\frac{3}{2},\frac{1}{2}\rangle \mapsto|1,0\rangle\\
	|\frac{3}{2},-\frac{1}{2}\rangle \mapsto|0,1\rangle\,\,\,\,
	|\frac{3}{2},-\frac{3}{2}\rangle \mapsto|0,0\rangle\,\,\,\text{.}\nonumber
\end{eqnarray}
The physical interpretation of our mapping is given in terms of the conserved quantities mentioned before. 
By interchanging the two sets of operators the number of flavors $L$ is replaced by the total spin $s$, while the total number of fermions is expressed by the spin-z projection.   
As an easy example of the application of our mapping, we start at $L=2$ with an already diagonal Hamiltonian which can only be built from combinations of number operators in the fermion representation:
\begin{equation}
	\hat{H}=E_1 \hat{n}_1 + E_2 \hat{n}_2\,\,\,\text{.}
\end{equation}
Here we assume the most simple form where only a linear combination of the number operators themselves appears instead of additional products between different number operators.
By expressing the fermionic operators in terms of spins with the help of our formula for the rewriting of diagonal operators (\ref{n_Sz_reversed}) and henceforth combining all terms with the respective same power of $\hat{S}_z$ operators we can transform this Hamiltonian to
\begin{eqnarray}\label{spin_hamiltonian}
	&\hat{H}
	=E_1 \hat{n}_1 + E_2 \hat{n}_2
	=E_1 p^{(2)}_1(\hat{S}_z) + E_2 p^{(2)}_2(\hat{S}_z)=\\
	&=E_1\left(\frac{1}{2} \mathds{1}_4 +\frac{13}{12} \hat{S}_z -\frac{1}{3} \hat{S}_z^3\right)+E_2\left(\frac{1}{2} \mathds{1}_4 -\frac{7}{6} \hat{S}_z +\frac{2}{3} \hat{S}_z^3\right)=\nonumber\\
	&=\frac{E_1 + E_2}{2}\mathds{1}_4 +\frac{13 E_1 -14 E_2}{12} \hat{S}_z +\frac{2E_2 -E_1}{3} \hat{S}_z^3\nonumber
\end{eqnarray}
If we treat a Hamiltonian pertaining to a three flavor system we get analogously by again using (\ref{n_Sz_reversed}) to get the replacements for the number operators:
\begin{eqnarray}\label{H_rewriting}
	&\hat{H}=\frac{E_1 + E_2+E_3}{2}\mathds{1}_8 +\frac{30251 E_1 -29774 E_2 
-34576E_3}{26880} \hat{S}_z+\\ 
	&+\frac{7(182E_2+496E_3 -215E_1)}{2880} \hat{S}_z^3 +\frac{61E_1 -34E_2 
-176E_3}{720}\hat{S}_z^5+\nonumber
\end{eqnarray}
\begin{eqnarray}\label{H_rewriting_end}
	&+\frac{-5E_1+2E_2+16E_3}{1260}\hat{S}_z^7\nonumber\\
	&\hat{n}_1=p^{(3)}_1(\hat{S}_z)=\frac{1}{2} \mathds{1}_8 +\frac{30251}{26880} 
\hat{S}_z -\frac{301}{576} \hat{S}_z^3+\frac{61}{720}\hat{S}_z^5 
-\frac{1}{252}\hat{S}_z^7\\
	&\hat{n}_2=p^{(3)}_2(\hat{S}_z)=\frac{1}{2} \mathds{1}_8 -\frac{14887}{13440} 
\hat{S}_z +\frac{637}{1440} \hat{S}_z^3-\frac{17}{360}\hat{S}_z^5 
+\frac{1}{630}\hat{S}_z^7\\
	&\hat{n}_3=p^{(3)}_3(\hat{S}_z)=\frac{1}{2} \mathds{1}_8 -\frac{2161}{1680} 
\hat{S}_z +\frac{217}{180} \hat{S}_z^3-\frac{11}{45}\hat{S}_z^5 
+\frac{4}{315}\hat{S}_z^7\,\,\,\text{.}
\end{eqnarray}

The rewriting (\ref{H_rewriting}) above can in principle be carried out for every $L$ analytically by using the general expressions (\ref{n_Sz_reversed}) of number operators in terms of the polynomials $p_\alpha^{(L)}(\hat{S}_z)$ but quickly yields very lengthy expressions, thus making it better suited for symbolic automated computations or general theoretical investigations.
\\
\\
An interesting consequence of these mappings is that it allows to obtain exact solutions for certain complicated fermionic Hamiltonians.
Consider for example the following Hamiltonian cubic in the fermion operators, which corresponds to the simple Hamiltonian of a system of a single spin precessing around a magnetic field $[b_{j}]$:
\begin{eqnarray}\label{cubicHam}
	&\hat{H}=\frac{1}{2}\left(2\sqrt{3}\hat{n}_1 \hat{c}_2^\dagger - 2\hat{c}_1^\dagger \hat{c}_2 -\sqrt{3} \hat{c}_2^\dagger\right)\bar{z_b}+\\
	&+\frac{1}{2}\left(2\sqrt{3}\hat{c}_2\hat{n}_{1} -2 \hat{c}_2^\dagger \hat{c}_1 -\sqrt{3}\hat{c}_2\right)z_b
	+b_z\left(\frac{-3}{2}\mathds{1}_4 + \hat{n}_2 +2\hat{n}_1\right)=\nonumber\\
	&=\frac{1}{2}\hat{S}_+ (b_x-ib_y) + \frac{1}{2} \hat{S}_-(b_x+ib_y)+b_z\hat{S}_z=\nonumber\\
	&=\frac{1}{2}(\hat{S}_x +i \hat{S}_y)(b_x -ib_y) + \frac{1}{2}(\hat{S}_x-i\hat{S}_y)(b_x+ib_y) + b_z\hat{S}_z=\nonumber\\
	&=\left[\hat{S}_x\,\,\,\hat{S}_y\,\,\,\hat{S}_z\right]
	\left[\begin{array}{c}
		b_x\\
		b_y\\
		b_z
	\end{array}\right]\,\,\,\text{.}\nonumber
\end{eqnarray}
Here the replacements (\ref{62}), (\ref{Szexplicit})

\begin{eqnarray}
	&\hat{S}_+=2\sqrt{3}\hat{n}_1 c_2^\dagger -2\hat{c}_1^\dagger \hat{c}_2 -\sqrt{3}\hat{c}_2^\dagger \\
	&\hat{S}_z=-\frac{3}{2}\mathds{1}_4 +\hat{n}_2 +2\hat{n}_1 
\end{eqnarray}
for the spin $s=3/2$ representation have been inserted as well as $z_b:=b_x+ib_y$.
If we apply a $\operatorname{SU}(2)$ transformation $U$ to the Hamiltonian in order to diagonalize it, we can make use of the fact that the adjoint action of the the operator $U$ on the vector operator $[\hat{S}_{j}]$ corresponds to a $\operatorname{SO}(3)$ rotation $R$ acting on the vector $[b_{j}]$ ( see e.g. \cite{boehm} for a mathematical explanation of this relation, although using index notation instead of the array notation used here):
\begin{eqnarray}
	U\hat{H}U^\dagger=
	\left[\hat{S}_x\,\,\,\hat{S}_y\,\,\,\hat{S}_z\right]
	R\left[\begin{array}{c}
		b_x\\
		b_y\\
		b_z
	\end{array}\right]\,\,\,\text{.}
\end{eqnarray}
By requiring that the rotated magnetic field vector only has a component along the new z-axis,
\begin{eqnarray}
	 R
	\left[\begin{array}{c}
			b_x\\
			b_y\\
			b_z
	\end{array}\right]=
	\left[\begin{array}{c}
			0\\
			0\\
			\tilde{b}_z
	\end{array}\right]\,\,\,\,\,\text{,}\,\,\,
	 b_x^2 + b_y^2 + b_z^2 = \tilde{b}_z^2
\end{eqnarray}
we carry out the diagonalization of the Hamiltonian so that only the $\hat{S}_z$ operator is included in it.
Since the spectrum $\sigma(\hat{H})$ of a Hamiltonian $\hat{H}$ is left unchanged by unitary transformations, we conclude that the spectrum of above given Hamiltonian is determined by the diagonal operator $\hat{S}_z$, scaled by $\tilde{b}_z$ as
\begin{eqnarray}
	\sigma(\hat{H})=\{\tilde{b}_z\frac{3}{2},\cdots, 
\tilde{b}_z\frac{-3}{2}\}\,\,\,\text{.}
\end{eqnarray}
If we replace the spin-z operator by its fermionic correspondence again we obtain the equality of spectra
\begin{eqnarray}
	&\sigma\left(\frac{1}{2}\left(2\sqrt{3}\hat{n}_1 \hat{c}_2^\dagger - 
2\hat{c}_1^\dagger \hat{c}_2 -\sqrt{3} \hat{c}_2^\dagger\right)\bar{z_b}
	+\frac{1}{2}\left(2\sqrt{3}\hat{c}_2\hat{n} -2 \hat{c}_2^\dagger \hat{c}_1 
-\sqrt{3}\hat{c}_2\right)z_b+\right.\nonumber\\
	&\left. +b_z\left(\frac{-3}{2}\mathds{1}_4 + \hat{n}_2 +2\hat{n}_1\right)\right)
	=\sigma\left(\tilde{b}_z\left(\frac{-3}{2}\mathds{1}_4 + \hat{n}_2 
+2\hat{n}_1\right)\right)\,\,\,\text{.}
\end{eqnarray}
Since this procedure relies on the same three dimensional rotation of the coefficient vector $[b_{j}]$ for any spin representation corresponding to ever more complicated appearing Hamiltonians when written in terms of fermionic operators, we can conclude, that the corresponding off-diagonal operators appearing can be associated purely with a rescaling of the diagonal part which in turn is comprised only of number operators.
The dependence on a three-dimensional vector as opposed to a vector of arbitrary dimension is of course rooted in the fact that the spin-algebra is a three-dimensional Lie-algebra.
Since the spin Hamiltonian from above example looks the same for any representation, we obtain a whole class of exactly solvable, but ever more complicated appearing fermionic Hamiltonians obtained by increasing the spin of the chosen representation.
Furthermore, our procedure of translating a Lie-algebra representation into a collection of fermionic operators relies essentially just on the construction of UODMs from fermionic operators since ladder operators of similar type are obtained for all semi-simple algebras for which the Cartan classification \cite{cartan1}, \cite{cartan2} is applicable.
Since the construction of UODMs is a problem we solved generally, it might also be feasible to make use of a translation of more complicated Lie-algebras than the spin algebra into collections of fermionic operators to obtain even more examples of highly non-trivial but still exactly solvable fermionic Hamiltonians, although the terms obtained by this might become too complicated for analytical treatments.

\subsection{Application to Spin Chains}
By making use of the same calculational methods as in (\ref{tech}) we can rewrite operators $\hat{S}^{(j)}_\alpha, j\in\{1,\cdots,P\},\alpha\in\{x,y,z,+,-\}$
for systems of $P$ spins situated at some lattice position, written as an upper index $j$, in terms of fermion operators of a sufficiently large amount of flavours $L$.
While a single spin operator in the spin $s$ representation is expressed by $L(s)=\operatorname{log}_2(2s+1)$ many fermion flavours we need $\operatorname{log}_2((2s+1)^P)$ many fermion flavours to express a spin operator pertaining to a system of $P$ spins.
Since a general theory of incorporating spin chains into our formalism is beyond the scope of this work, we restrict ourselves to a simple example of a common Ising interaction term of the form
\begin{eqnarray}\label{ising}
	\hat{S}^{(1)}_z\hat{S}^{(2)}_z
\end{eqnarray}
for the case of just two interacting spins in the $s=3/2$ representation.
By means of (\ref{Szexplicit}) we can express the spin operators by number operators as
\begin{eqnarray}\label{SzP2}
	&\hat{S}^{(1)}_z:=\hat{S}_z\otimes\mathds{1}_4=\left(-\frac{3}{2}\mathds{1}_4+\hat{n}_2(2) +2\hat{n}_1(2)\right)\otimes\mathds{1}_4=\\
	&=-\frac{3}{2}\mathds{1}_{16}+\hat{n}_2(4)+2\hat{n}_1(4)\nonumber
\end{eqnarray}
\begin{eqnarray}\label{SzP2_2}
	&\hat{S}^{(2)}_z:=\mathds{1}_4\otimes\hat{S}_z=\mathds{1}_4\otimes\left(-\frac{3}{2}\mathds{1}_4+\hat{n}_2(2) +2\hat{n}_1(2)\right)=\\
	&=-\frac{3}{2}\mathds{1}_{16}+\hat{n}_4(4)+2\hat{n}_3(4)\nonumber
\end{eqnarray}
where we denoted the number of flavours explicitly for the number operators as $\hat{n}_\alpha(L)$.
Inserting (\ref{SzP2}) and (\ref{SzP2_2}) in (\ref{ising}) gives us the number operator representation of the Ising interaction as
\begin{eqnarray}
	&\hat{S}^{(1)}_z\hat{S}^{(2)}_z=\frac{9}{4}\mathds{1}_{16} -\frac{3}{2}\left(2\hat{n}_1(4) +\hat{n}_2(4)+2\hat{n}_3(4)+\hat{n}_4(4)\right)+\\
	&+\hat{n}_2(4)\hat{n}_4(4) +2\hat{n}_2(4)\hat{n}_3(4) +2\hat{n}_1(4)\hat{n}_4(4) +4\hat{n}_1(4)\hat{n}_3(4)\,\,\,\text{.}\nonumber
\end{eqnarray}

\newpage
\section{Conclusion}
We have constructed a new spin fermion mapping that can be used to express a system of $L$ fermionic operators in terms of a single spin algebra in the $s(L)=(2^L -1)/2$ representation of $\mathfrak{su}(2)$.
An analytical form for the $\hat{S}_z$ operator in terms of fermionic number operators and reversely is obtained by making use of the properties of Vandermonde matrices.
Since our formalism is built on representing a general UODM, it might also serve to rewrite other observable algebras than $\mathfrak{su}(2)$ in terms of fermionic operators inside of the same representation space.
The formalism presented is well suited to be implemented for numerical or symbolic computations for larger dimensions, but is founded on analytical and exact computations, which makes it useful for specific analytical calculations requiring a single replacement of one set of operators by the other.
Inconveniences that still remain, are the complexity of the obtained expansions as well as the recursively defined basis transformation matrix $V_{(\text{c})}^{-1}(L)$ of the fermion basis and the requirement of taking matrix roots to obtain the component vectors of the fermion operators, where the latter could have been prevented by allowing more general vector spaces of matrices than just the UODMs.
For example, one could also use $E^{(S)}_\alpha(s)=\hat{S}_+^\alpha \hat{S}_z^\beta\hat{S}_-^\alpha$ as the spin operator basis, which we decided against in order to keep the formalism simpler.
Our methodology of coupling several systems to form a single large spin is similar, but not equivalent, to usual spin-coupling theory involving the Clebsch-Gordan coefficients, whose role roughly is taken by the expansion coefficients $(V^T_{(c)}(L))^{-1}[x_{j}], (V^T_{(S)}(s))^{-1}[x_j]$ for the UODM operators (\ref{general_expansion}) and $\tilde{y}^{(\alpha)}$ (\ref{y_beta}) for the diagonal number operators, while the $\hat{S}_z$ operators are given by (\ref{Szexplicit}) without the need to determine coefficients.
Since for usual coupling theory there have been semi-classical treatments made available \cite{Littlejohn1},\cite{Littlejohn2},\cite{Littlejohn3} that express the coupling coefficients by classical quantities, it might be feasible to also apply semi-classical approximations to our formalism in order to counteract the substantially growing complexity of the obtained operator-expansions with increasing $L$.
In this context we note that, due to the exponentiation $2^L$ in the dimension of the Hilbert space, it might be possible to relate already a system of as few as eight fermions to some form of semi-classical treatment, because this number shifts the corresponding spin representation $s=255/2$ to the semi-classical domain starting roughly around $s=100$ for common systems of a single spin, like the Kicked Top \cite{KT},\cite{KT_SC}.
Further investigations on this should be to identify the right classical system with such a system of $L\geq 8$ fermion flavors.
This might be achieved by either using the corresponding spin-Hamiltonian directly, which might become unpractical analytically, since it involves $\approx 10^2$ different powers of spins operators or to truncate it in some well defined manner and then use the usual $[\hat{S}_{j}/(s+\frac{1}{2})]\to [\sqrt{1-p^2}\cos(q),\sqrt{1-p^2}\sin(q),p]^T$ replacement in terms of symplectic Bloch-sphere coordinates $q,p$.
Other novel approaches not directly related to our formalism but to the investigation of semi-classical treatments of fermionic systems can be found in \cite{Fermion_Integrability_Regensburg},\cite{Spin_Echo_Regensburg},\cite{Regensburg_sc_propagator}.
Regarding systems of multiple spins, like spin chains, we show that the tools developed in this work can be applied to the operators describing these systems as well, although the complexity of the corresponding fermionic rewritings increases much more quickly since the number of fermion flavours needed to express a spin operator of a $P$ spin system is $\operatorname{log}_2((2s+1)^P)$.
An interesting question for future works would be what spin chain systems can be mapped onto fermionic systems linear in the fermionic number operators and hence of the type described by Hamiltonians as in the first half of section 3.1.

\appendix
\section{Proof of CAR Fulfillment}\label{A.1.}
We want to show, that the above defined operators obey the given CAR.
Assume $\alpha\leq\beta$, then re-partition the factors in the definition of $\hat{c}_\alpha$ so that tensor-product powers of equal length can be compared. 
Also the fact
 \begin{equation}
  \{A\otimes B,C\otimes D\}=
  AC\otimes BD + CA\otimes DB\stackrel{{[B,D]} {=0}}
  =\{A,C\}\otimes BD
 \end{equation}
and equivalently
 \begin{equation}
  \{A\otimes B,C\otimes D\}\stackrel{{[A,C]}{=0}}=AC\otimes\{B,D\}
 \end{equation}
will be used in near the end of the proof.
\begin{eqnarray}
 &\{\hat{c}_\alpha,\hat{c}^\dagger_\beta\}=
 \{r^{\otimes(\alpha-1)}\otimes\hat{c}\otimes \mathds{1}_2^{\otimes(L-\alpha)},
 (r^\dagger)^{\otimes(\beta-1)}\otimes \hat{c}^\dagger \otimes 
\mathds{1}_2^{\otimes(L-\beta)}\}\\
 &=\{r^{\otimes(\alpha-1)}\otimes\hat{c}\otimes 
\mathds{1}_2^{\otimes(\beta-\alpha)}\otimes \mathds{1}_2^{\otimes(L-\beta)},
 (r^\dagger)^{\otimes(\alpha-1)}\otimes (r^\dagger)^{\otimes(\beta-\alpha)}\otimes 
\hat{c}^\dagger \otimes \mathds{1}_2^{\otimes(L-\beta)}\}\nonumber\\
 &=r^{\otimes(\alpha-1)} (r^\dagger)^{\otimes(\alpha-1)}\otimes\{c\otimes 
\mathds{1}_2^{\otimes(\beta-\alpha)},
(r^\dagger)^{\otimes(\beta-\alpha)}\otimes 
\hat{c}^\dagger\}\otimes\mathds{1}_2^{\otimes(L-\beta)}\mathds{1}_2^{\otimes(L-\beta)
}\nonumber\\
 &=\mathds{1}_2^{\otimes(\alpha-1)}\otimes\{\hat{c}\otimes 
\mathds{1}_2^{\otimes(\beta-\alpha)},
  (r^\dagger)^{\otimes(\beta-\alpha)}\otimes 
\hat{c}^\dagger\}\otimes\mathds{1}_2^{\otimes(L-\beta)}\nonumber
\end{eqnarray}
Now consider just the anti-commutator in the middle.
We have to check the cases $\alpha=\beta$ and $\alpha <\beta$. 
For $\alpha=\beta$ this anti-commutator yields just
\begin{equation}
 \{\hat{c}\otimes \mathds{1}_2^{\otimes(0)},
 (r^\dagger)^{\otimes(0)}\otimes \hat{c}^\dagger\}=\{c,\hat{c}^\dagger\}=\mathds{1}_2
\end{equation}
and for $\alpha<\beta$
\begin{equation}
  \{\hat{c}\otimes \mathds{1}_2^{\otimes(\beta-\alpha)},
 (r^\dagger)^{\otimes(\beta-\alpha)}\otimes \hat{c}^\dagger\}=
\end{equation}
\begin{equation*}
 =\{\hat{c},r^\dagger\}\otimes \mathds{1}_2^{\otimes (\beta-\alpha)} 
(r^{\otimes(\beta-\alpha-1)}\otimes \hat{c}^\dagger)
\end{equation*}
The first case, together with the left over tensor-powers of identity matrices from above gives the correct relations
\begin{equation}
\{\hat{c}_\alpha,\hat{c}^\dagger_\alpha\}=\delta_{\alpha\alpha}\mathds{1}_{2^L}
=\mathds{1}_{2^L}
\end{equation}
while we can make the anti-commutator vanish for the second case, for all the different possibilities of combined ladder operators listed above by demanding
\begin{equation}\label{A11}
\{\hat{c},r\}=0\hspace{1cm}\{\hat{c},r^\dagger\}=0\hspace{1cm}\{\hat{c}^\dagger,
r^\dagger\}=0
\end{equation}
so that the first term in the middle anti-commutator vanishes. 
The only solution of these equations (\ref{A11}) is given by $\sigma_z$.
Of course the case $\alpha>\beta$ can be obtained simply by relabeling all terms and therefore needs no further consideration. This completes the proof.

\section{Linear Independence of Fermionic Basis $E^{(c)}_\alpha(L+1)$}\label{A.2.}
The linear independence of the basis operators $E^{(c)}_\alpha(L+1)$ can be proven inductively by assuming the linear independence of the $E^{(c)}_\alpha(L)$ and then showing, that the procedure for increasing $L\rightarrow L+1$ preserves it.
As before we have the set of basis operators split into three parts $\{E^{(c)}_\alpha(L)\}_{\alpha\in \{1,\cdots , 2^{L-1}-1\}}\cup\{\hat{c}_1^\dagger \hat{c}_2\cdots \hat{c}_L\}\cup\{E^{(c)}_\alpha(L)\}_{\alpha\{\in 2^{L-1}+1,\cdots,2^L-1\}}$ for which we temporarily want to use the notations $X_\alpha(L),Y(L),Z_\alpha(L)$ for elements of the respective three subsets.
We assume, that all of the basis operators for the dimensional parameter $L$ are linearly independent.
Since the elements of each part $X_\alpha(L),Y(L),Z_\alpha(L)$ obviously do not lose their linear independence by the operations (\ref{recursion_relations})-(\ref{recursion_relations3}) given above, the new basis operators are already guaranteed to be linearly independent within each of these three subsets so that we only have to check for possible linear dependence between different subsets.
For this we attempt a proof by contradiction by assuming, that $X_\alpha(L+1)$ and $Z_\alpha(L+1)$ are linearly dependent, so that we have an expansion with real coefficients $A_{\alpha\beta}$ where the tensor-product from the left with $\sigma_z$ or $\mathds{1}_2$  corresponds respectively to the index-shift operation for basis elements that contain an odd or even amount of products of two-dimensional operators.
We only show the proof for the former case of basis elements containing an odd number of products on the right hand side of the equation, the other case can be seen to be treatable exactly analogously by exchanging the $\sigma_t$ on the right side by 
$\mathds{1}_2$:
\begin{eqnarray}
	\sum\limits_{\beta=1}^{2^L -1} A_{\alpha\beta} \hat{n}\otimes 
E^{(c)}_\beta(L)=Z_\alpha(L+1)\\
	\Leftrightarrow 
	\sum\limits_{\beta=1}^{2^L -1} A_{\alpha\beta} \frac{\sigma_z + \mathds{1}_2}{2} 
\otimes E^{(c)}_\beta(L)=\sigma_z\otimes E^{(c)}_\alpha(L)\\
	\text{choose }A_{\alpha\alpha}=2\\
	\Leftrightarrow \sum\limits_{\beta=1}^{2^L -1} A_{\alpha\beta} 
\frac{\mathds{1}_2}{2} \otimes 
E^{(c)}_\beta(L)+\sum\limits_{\substack{\beta=1\\\beta\neq\alpha}}^{2^L -1} 
A_{\alpha\beta} \frac{\sigma_z}{2} \otimes E^{(c)}_\beta(L)=0
\end{eqnarray}
Since the two sums in the last step have a different first tensor factor, but identical second factor, they have to vanish individually.
Since the basis element $Y(L)$ has only a single non-zero element for every $L$,
\begin{eqnarray}
	Y(L)=\left[\begin{array}{ccc}
					\mathbf{0} & 0 & \mathbf{0}\\
					\vdots & (-1)^{L-1} & \vdots\\
					\mathbf{0} & 0 & \mathbf{0}
					\end{array}\right]
\end{eqnarray}
and since the basis elements $X_\alpha(L)$, $Z_\alpha(L)$ only have entries on the blocks below or above the middle matrix row separating the blocks, it is immediately clear that $Y(L)$ is linearly independent from the others.
This gives a contradiction to the assumption, that $E^{(c)}_\alpha(L+1)$ are linearly independent and hence completes the proof.

\section{Explicit Construction of Fermionic Basis}\label{A.3.}
We start with the already motivated basis for $L=2$, $\{\hat{c}_1^\dagger \hat{c}_1 \hat{c}_2^\dagger, \hat{c}_1^\dagger \hat{c}_2 , \hat{c}_2^\dagger\}$ and use the recursion relations (\ref{recursion_relations}) to obtain the basis for $L=3$:
\begin{equation}
	\{E^{(c)}_1(3),E^{(c)}_2(3),E^{(c)}_3(3)\}=\hat{n}_1 \text{index-shift}(\{\hat{c}_1^\dagger \hat{c}_1 \hat{c}_2^\dagger, \hat{c}_1^\dagger \hat{c}_2 , \hat{c}_2^\dagger\})=
\end{equation}
\begin{equation*}
	=\{\hat{n}_1 \hat{c}_2^\dagger \hat{c}_2 \hat{c}_3^\dagger, \hat{n}_1 \hat{c}_2^\dagger \hat{c}_3, \hat{n}_1 \hat{c}_3^\dagger \}=\{\hat{n}_1 \hat{n}_2 \hat{c}_3^\dagger, \hat{n}_1 \hat{c}_2^\dagger \hat{c}_3, \hat{n}_1 \hat{c}_3^\dagger\}
\end{equation*}
\begin{equation}
	\{E^{(c)}_5(3),E^{(c)}_6(3),E^{(c)}_7(3)\}=\text{index-shift}(\{\hat{c}_1^\dagger \hat{c}_1 \hat{c}_2^\dagger, \hat{c}_1^\dagger \hat{c}_2 , \hat{c}_2^\dagger\})=
\end{equation}
\begin{equation*}
	=\{\hat{c}_2^\dagger \hat{c}_2 \hat{c}_3^\dagger, \hat{c}_2^\dagger \hat{c}_3,  \hat{c}_3^\dagger \}=\{ \hat{n}_2 \hat{c}_3^\dagger,  \hat{c}_2^\dagger \hat{c}_3, \hat{c}_3^\dagger\}
\end{equation*}
\begin{equation}
	E^{(c)}_4(3)=\hat{c}_1^\dagger \hat{c}_2 \hat{c}_3
\end{equation}
For the just stated basis this gives
\begin{equation}\label{appendix_V}
	(V^{\pm 1}_{(\text{c})}(L=3))^{T}=\left[\begin{array}
				{ccccccc}
				1&0&1&0&1&0&1\\
				0&-1&0&0&0&\pm -1&0\\
				0&0&-1&0&0&0&-1\\
				0&0&0&1&0&0&0\\
				0&0&0&0&-1&0&-1\\
				0&0&0&0&0&-1&0\\
				0&0&0&0&0&0&1
			\end{array}\right]\,\,\,\text{,}
\end{equation}
where the $\pm$ in the exponent means either the matrix itself (+) or its inverse (-), corresponding to the same order of signs in the single multivalued entry of the array on the right hand side of the equation (\ref{appendix_V}).

\section{Explicit Construction of Spin operators from Fermionic Operators for $L=3$ and $L=4$}\label{A.4.}
We use the fermionic basis computed in the previous appendix combined with the formula for the components of the spin operator (\ref{Sp_components}) and combine them as in (\ref{transpose_operation}) to obtain:
Spin $7/2$ corresponding to $L=3$:
\begin{eqnarray}\label{spin72exp}
	&\hat{S}_+=2(\sqrt{7}+\sqrt{15})\hat{n}_1 \hat{n}_2 \hat{c}_3^\dagger-(\sqrt{7}+\sqrt{15})\hat{n}_1 \hat{c}_3^\dagger\\
	&+4 \hat{c}_1^\dagger \hat{c}_2 \hat{c}_3 -(\sqrt{7}+\sqrt{15})\hat{n}_2\hat{c}_3^\dagger -2\sqrt{3} \hat{c}_2^\dagger \hat{c}_3 +\sqrt{7}\hat{c}_3^\dagger\nonumber\\
	&\hat{S}_+\in\operatorname{Mat}(8,8)\hspace{1.5cm} \hat{c}_\alpha\in\operatorname{Mat}(8,8)\nonumber
\end{eqnarray}
In the same way we obtain the expansion for four fermion flavours:
Spin $15/2$ corresponding to $L=4$:
\begin{eqnarray}\label{spin72exp}
	&\hat{S}_+=(2\sqrt{15}+2\sqrt{39}+2\sqrt{55}+6\sqrt{7})\hat{n}_1\hat{n}_2\hat{n}_3 \hat{c}_4^\dagger\\
	&+(-4\sqrt{7}+4\sqrt{15})\hat{n}_1\hat{n}_2\hat{c}_3^\dagger \hat{c}_4 -
	(3\sqrt{7}+\sqrt{55}+\sqrt{39}+\sqrt{15}) \hat{n}_1 \hat{n}_2 \hat{c}_4^\dagger\nonumber\\
	&+8\sqrt{3} \hat{n}_1 \hat{n}_2 \hat{c}_4^\dagger-
	(3\sqrt{7} +\sqrt{55} +\sqrt{39}\sqrt{15})\hat{n}_1 \hat{n}_3 \hat{c}_4^\dagger\nonumber\\
	&+(2\sqrt{7}-2\sqrt{15})\hat{n}_1 \hat{c}_3^\dagger \hat{c}_4
	+(\sqrt{15}+3\sqrt{7})\hat{n}_1 \hat{c}_4^\dagger\nonumber\\
	&-8 \hat{c}_1^\dagger \hat{c}_2 \hat{c}_3 \hat{c}_4
	-(3\sqrt{7} +\sqrt{55}+\sqrt{39}+\sqrt{15})\hat{n}_2 \hat{n}_3 \hat{c}_4^\dagger\nonumber\\
	&+(2\sqrt{7}-2\sqrt{15}) \hat{n}_2 \hat{c}_3^\dagger \hat{c}_4
	+(\sqrt{15}+\sqrt{55}) \hat{n}_2 \hat{c}_4^\dagger\nonumber\\
	&-4\sqrt{3} \hat{c}_2^\dagger \hat{c}_3 \hat{c}_4
	+(\sqrt{15}+\sqrt{39}) \hat{n}_3 \hat{c}_4^\dagger
	-2\sqrt{7} \hat{c}_3^\dagger \hat{c}_4
	-\sqrt{15} \hat{c}_4^\dagger\nonumber\\
	&\hat{S}_+\in\operatorname{Mat}(16,16)\hspace{1.5cm} \hat{c}_\alpha\in\operatorname{Mat}(16,16)
\end{eqnarray}
\section{Explicit Construction of Fermionic Operators from Spin Operators for $L=3$}\label{A.5.}
Because of the lengthy expressions inside of the matrix $V_{(S)}^{-1}$ (\ref{def_Vs}) we only give the result of the multiplication with the coefficient vectors.
The basis operators $E^{(S)}_j(s)$ were defined in (\ref{spin_basis}) and for taking the matrix roots of the fermionic operators we used the technique from section 2.5 which resulted in (\ref{ferm_comp})-(\ref{ferm_comp_end}) and which is only one of multiple possible solutions for taking the matrix root.
\begin{eqnarray}
	&\hat{c}_{\alpha}^\dagger=\left([E^{(S)}_1\left(\frac{7}{2}\right) \cdots E^{(S)}_7\left(\frac{7}{2}\right)] (V_{(S)}^{-1}(3))^{T}[x^{(c,\alpha)}_j]\right)^{2^{3-\alpha}}=\\
	 & =\left([E^{(S)}_1\left(\frac{7}{2}\right) \cdots E^{(S)}_7\left(\frac{7}{2}\right)][\tilde{x}_j^{(c,\alpha)}]\right)^{2^{3-\alpha}}\\
	& [\tilde{x}_j^{(c,1)}]=\left[\begin{array}
									{c}
									\frac{-7\sqrt{3}+35\sqrt{15}}{1536}+\frac{\sqrt{7}}{3584}+\frac{175}{1024}\\
									\frac{75}{256}+\frac{-67\sqrt{3}}{5760}+\frac{3\sqrt{7}}{4480}+\frac{95\sqrt{15}}{1152}\\
									\frac{-463}{2304}+\frac{41\sqrt{7}}{40320}+\frac{9\sqrt{3}+31\sqrt{15}}{640}\\
									\frac{-47\sqrt{15}}{720}+\frac{-\sqrt{7}}{336}+\frac{7\sqrt{3}}{144}+\frac{17}{96}\\
									\frac{-37\sqrt{15}}{1440}+\frac{-\sqrt{7}}{2016}+\frac{5\sqrt{3}}{288}+\frac{41}{576}\\
									\frac{-1}{48}+\frac{\sqrt{15}-\sqrt{3}}{120}+\frac{\sqrt{7}}{840}\\
									\frac{-1}{144}+\frac{\sqrt{15}-\sqrt{3}}{360}+\frac{\sqrt{7}}{2520}
							 \end{array}\right]\\
 	& [\tilde{x}_j^{(c,2)}]=\left[\begin{array}
									{c}
									\frac{-7\sqrt{3}}{1536}+\frac{3\sqrt{7}}{1792}+\frac{35\sqrt{15}}{768}\\
									\frac{67\sqrt{3}}{5760}+\frac{3\sqrt{7}}{4480}+\frac{85\sqrt{15}}{1152}\\
									\frac{-73\sqrt{15}}{1440}+\frac{-5\sqrt{7}}{672}+\frac{9\sqrt{3}}{640}\\
									\frac{-7\sqrt{15}}{240}+\frac{-\sqrt{7}}{336}+\frac{7\sqrt{3}}{144}\\
									\frac{\sqrt{7}}{336}+\frac{\sqrt{15}}{144}+\frac{5\sqrt{3}}{288}\\
									\frac{-\sqrt{3}}{120}+\frac{\sqrt{7}}{840}+\frac{\sqrt{15}}{360}\\
									\frac{-\sqrt{3}}{360}\\
							 \end{array}\right]\\
& [\tilde{x}_j^{(c,3)}]=\left[\begin{array}
							{c}
							\frac{-35\sqrt{15}}{1536}+\frac{\sqrt{7}}{3584}\\
							\frac{-95\sqrt{15}}{1152}+\frac{3\sqrt{7}}{4480}\\
							\frac{-31\sqrt{15}}{640}+\frac{-41\sqrt{7}}{40320}\\
							\frac{-\sqrt{7}}{336}+\frac{47\sqrt{15}}{720}\\
							\frac{-\sqrt{7}}{2016}+\frac{37\sqrt{15}}{1440}\\
							\frac{-\sqrt{15}}{120}+\frac{\sqrt{7}}{840}\\
							\frac{-\sqrt{15}}{360}+\frac{\sqrt{7}}{2520}\\
					 \end{array}\right]
\end{eqnarray}

\section{Solution of Recursion Relation: Example}\label{A.6.}
An example for the pattern-matrix for up to $L=4$ we get:
\begin{eqnarray}
	\\
    &\mathcal{P}_4=\left[\begin{array}
						{ccc}
                      1 & 1 & 1\\
                      1 & 1 & 0\\
                      1 & 1 & -1\\
                      1 & 0 & 0\\
                      1 & -1& 1\\
                      1 & -1& 0\\
                      1 & -1& -1\\
                      0 & 0 & 0\\
                      -1 & 1&1 \\
                      -1 &1&0\\
                      -1 &1&-1\\
                      -1 &0&0\\
                      -1 &-1&1\\
                      -1 &-1&0\\
                      -1 &-1&-1\\
                  \end{array}\right]=[r_3\,\,\, r_2\oplus 0\oplus r_2\,\,\, r_1 \oplus 0 \oplus r_1 \oplus 0\oplus r_1\oplus 0 \oplus r_1]\nonumber\\
				  \\
    &\mathcal{P}_3=\left[\begin{array}
					{cc}
                  1 & 1\\
                  1 & 0\\
                  1 & -1\\
                  0 & 0\\
                  -1 & 1\\
                  -1 & 0\\
                  -1 & -1
                 \end{array}\right]=[r_2\,\,\,r_1\oplus 0 \oplus r_1]\\
   &\mathcal{P}_2=\left[\begin{array}
   					{c}
                  1\\
                  0\\
                 -1
                  \end{array}\right]=r_1
\end{eqnarray}
As a further example we apply the above formula to compute $D_2$, where $2^L -1 = 2^2 -1 =3$ summands appear:
\begin{eqnarray}
    &\Pi_1=\mathcal{O}^{(\mathcal{P}_{2,1})_1}=\mathcal{O}^{(1)}=\mathds{P}_{\uparrow}\\        
    &\Pi_2=\mathcal{O}^{(\mathcal{P}_{2,1})_2}=\mathcal{O}^{(0)}=1\nonumber \\
    &\Pi_3=\mathcal{O}^{(\mathcal{P}_{2,1})_3}=\mathcal{O}^{(-1)}=\mathds{P}_{\downarrow}\nonumber\\
    &m_1=1-\log_2(\dim(\Pi_1))=1-1=0\\ 
    &m_2=1-\log_2(\dim_1(\Pi_2))=1-0=1\nonumber\\ 
    &m_3=1-\log_2(\dim_1(\Pi_3))=1-1=0\nonumber\\
    &\Rightarrow D_2=D_2([x_{j}])=x_1\Pi_1 \otimes \hat{c}^\dagger \otimes \hat{c}^{\otimes m_1} + 
    x_2\Pi_2 \otimes \hat{c}^\dagger \otimes \hat{c}^{\otimes m_2}+
    x_3\Pi_3 \otimes \hat{c}^\dagger \otimes \hat{c}^{\otimes m_3}=\nonumber\\
	\\
    &= x_1 \mathds{P}_{\uparrow} \otimes \hat{c}^{\dagger} \otimes \hat{c}^{\otimes 0} + x_2 1\otimes \hat{c}^\dagger \otimes \hat{c}^{\otimes 1} + x_3 \mathds{P}_{\downarrow} \otimes \hat{c}^\dagger \otimes \hat{c}^{\otimes 0}\nonumber
\end{eqnarray}
Here we make use of $A^{\otimes 0}=1\in\mathbb{R}$ and $1\otimes B=B$ for any matrices $A$, $B$.
\\
\\
\textbf{Acknowledgments}\\
We gratefully acknowledge fruitful discussions with W. Hogger and J.D. Urbina at the Dreiburg conference on Many-Body Quantum Chaos, Duisburg 2019.
One of us (FM) thanks the German Research Foundation (DFG) for funding within the project \textit{Semiclassical Approach to Spectral Statistics of Interacting Quantum Many-Body Systems} (project number 402552305).


\section*{References}
\begin{thebibliography}{10}
\bibitem {JWT}P.\ Jordan, E.\ Wigner, \textit{Z.\ Phys.} \textbf{47}, 631 (1928).
\bibitem {General_JWT_Batista}C.\ D. \  Batista , G. \ Ortiz \textit{Phys.\ Rev.\ Lett.} \textbf{86}, 1082 (2001).
\bibitem {General_JWT_Kochmanski}M.\ S.\ Kochma$\acute{\operatorname{n}}$ski, \textit{J.\ Exp.\ Theor.\ Phys.} \textbf{84}, 940 (1997).
\bibitem {General_JWT_Galitski} V.\ Galitski, \textit{Phys.\ Rev.\ B} \textbf{82}, 060411 (2010).
\bibitem {Dobrov}S.\ V.\ Dobrov, \textit{J.  Phys. A: Math. Gen.} \textbf{36}, L503 (2003).
\bibitem{minami5} K.\ Minami, \textit{J.\ Phys.\ A:\ Math.\ Gen.} \textbf{29}, 6395 (1996).
\bibitem{minami6} K.\ Minami, \textit{J.\ Phys.\ Soc.\ Jpn.} \textbf{67}, 2255 (1998).
\bibitem{Guhr_quelle} P.J. Brussard, P. W. M. Glaudemans, \textit{Shell Model Applications in Nuclear Spectroscopy}, North Holland Publishing Comp (1977).
\bibitem{CLEBSCH_GORDAN} G.\ Auletta, M.\ Fortunato, G\ Parisi, \textit{Quantum Mechanics}, Cambridge University Press (2009).
\bibitem {Miller1} W.\ H.\ Miller, K.\ A.\ White, \textit{J. Chem. Phys.} \textbf{84}, 5059 (1986).
\bibitem {Miller2} H.\ D.\ Meyer, W.\ H.\ Miller, \textit{J.\ Chem.\ Phys.} \textbf{71}, 2156 (1979).
\bibitem {Miller3} C.\ W.\ Mcmurdy, H.\ D.\ Meyer, W.\ H.\ Miller, \textit{J.\ Chem.\ Phys.} \textbf{70}, 3177 (1979).
\bibitem {MINIMAL_POLYNOMIAL}M.\ T.\ Nair, A.\ Singh, \textit{Linear Algebra}, Springer Singapore (2018).
\bibitem {VANDERMONDE_INVERSE} E. A. Rawashdeh, \textit{Mat.\ Vesn.} \textbf{71}, 207 (2019).
\bibitem {VANDERMONDE_INVERSE2_AND_DEF} L.\ R.\ Turner, \textit{NASA\ TN\ D}-3547 (1966).
\bibitem {ESP} R.\ P.\ Stanley, \textit{Enumerative Combinatorics Vol.2}, Cambridge University Press (1999).
\bibitem{boehm} M. Böhm, \textit{Lie-Gruppen und Lie-Algebren in der Physik}, Springer-Verlag Berlin Heidelberg (2011).
\bibitem{cartan1} F.\ Iachello, \textit{Lie Algebras and Applications}, Second Edition, Springer-Verlag Berlin Heidelberg (2015).
\bibitem{cartan2} P.\ Di\ Franchesco, P.\ Mathieu, D.\ S$\acute{\operatorname{e}}$n$\acute{\operatorname{e}}$chal, \textit{Conformal Field Theory}, Springer-Verlag New York (1997).
\bibitem {Littlejohn1} R.\ G.\ Littlejohn, L.\ Yu, \textit{J.\ Phys.\ Chem.\ A} \textbf{113}, 14904 (2009).
\bibitem {Littlejohn2} H.\ M.\ Haggard, R.\ G.\ Littlejohn, \textit{Class.\ Quant.\ Grav.} \textbf{27}, 135010 (2010).
\bibitem {Littlejohn3} L.\ Yu, R.\ G.\ Littlejohn, \textit{Phys.\ Rev.\ A} \textbf{83}, 052114 (2011).
\bibitem {KT} F.\ Haake, M.\ Kus, R.\ Scharf, \textit{Z.\ Phys.\ B} \textbf{65}, 381 (1987).
\bibitem {KT_SC} M.\ Kus, F.\ Haake, B.\ Eckhardt, \textit{Z.\ Phys.\ B} \textbf{92}, 221 (1993).
\bibitem {Fermion_Integrability_Regensburg}S.\ Grosse-Holz, T.\ Engl,\ K.\ Richter,\ J.\ Urbina, \textit{Acta\ Phys.\ Pol.} \textbf{128}, 994 (2015).
\bibitem {Spin_Echo_Regensburg}T.\ Engl, J.\ Urbina,K.\ Richter, P.\ Schlagheck, \textit{Phys.\ Rev.\ A} \textbf{98}, 013630 (2018).
\bibitem {Regensburg_sc_propagator} T.\ Engl, P.\ Plössl, J.\ D.\ Urbina, K.\ Richter, \textit{Theor.\ Chem.\ Acc.} \textbf{133}, 1563 (2014).
\end{thebibliography}
\end{document}